\documentclass[5p,authoryear]{elsarticle}
\usepackage{amsmath}
\usepackage{caption}
\usepackage{color}
\usepackage{graphicx}
\usepackage{listings}
\usepackage{siunitx}
\usepackage[export]{adjustbox}
\usepackage[normalem]{ulem}
\usepackage[many]{tcolorbox}
\usepackage[colorlinks=true,breaklinks,urlcolor=blue,citecolor=red,linkcolor=red,bookmarks=true]{hyperref}
\usepackage{hyphenat}

\journal{Astronomy and Computing}
\graphicspath{ {png/} }

\lstdefinelanguage{sqlmore}{                                                          
    language = sql,
    morekeywords = {language
                   ,import
                   ,function
                   ,double
                   ,python
                   ,return
                   ,returns
                   ,declare
                   ,sqrt
                   ,sin
                   ,cos
                   ,asin
                   ,acos
                   ,log10
                   ,pi
                   ,floor
                   ,alpha
                   ,radians
                   ,degrees}
}

\newtcolorbox{cross}{blank,breakable,parbox=false,
  overlay={\draw[red,line width=5pt] (interior.south west)--(interior.north east);
    \draw[red,line width=5pt] (interior.north west)--(interior.south east);}}

\begin{document}

\begin{frontmatter}

\title{Fast in-database cross-matching of high-cadence, high-density source lists with an up-to-date sky model}

\author[cwi]{Bart Scheers\corref{cor1}}
\ead{bartscheers@gmail.nl}
\author[ru,nova]{Steven Bloemen}
\author[cwi]{Hannes M\"uhleisen}
\author[princ,ru]{Pim Schellart}
\author[leiden]{Arjen van Elteren}
\author[cwi]{Martin~Kersten}
\author[ru]{Paul~J.~Groot}

\cortext[cor1]{Corresponding author}

\address[cwi]{CWI -- Centrum Wiskunde \& Informatica, 
              PO Box 94079, 1090 GB Amsterdam, The Netherlands}
\address[ru]{Department of Astrophysics, IMAPP, Radboud University, 
             6500 GL Nijmegen, The Netherlands}
\address[nova]{NOVA Optical InfraRed Instrumentation Group, 
               Oude Hoogeveensedijk 4, 7991 PD Dwingeloo, The Netherlands}
\address[princ]{Department of Astrophysical Sciences, Princeton University,
                Princeton, NJ 08544, USA}
\address[leiden]{Leiden Observatory, Leiden University, 
                 PO Box 9513, 2300 RA Leiden, The Netherlands}

\begin{abstract}

Coming high-cadence wide-field optical telescopes 
will image hundreds of thousands of sources per minute.
Besides inspecting the near real-time data streams for 
transient and variability events,
the accumulated data archive is a wealthy laboratory for making 
complementary scientific discoveries.

The goal of this work is to 
optimise column-oriented database techniques
to enable the construction of a full-source and light-curve database 
for large-scale surveys,
that is accessible by the astronomical community.

We adopted LOFAR's Transients Pipeline as the baseline 
and modified it to enable the processing of optical images
that have much higher source densities.
The pipeline adds new source lists to the archive database, 
while cross-matching them with the known cataloged sources 
in order to build a full light-curve archive.
We investigated several techniques of indexing and partitioning the largest tables,
allowing for faster positional source look-ups in the cross matching algorithms.
We monitored all query run times 
in long-term pipeline runs
where we processed a subset of IPHAS data
that have image source density peaks over 
$170,000$ per field of view
($500,000$\,deg$^{-2}$).

Our analysis demonstrates that horizontal table partitions of declination widths 
of one-degree control 
the query run times.
Usage of 
an
index strategy where the partitions are densily sorted according to
source declination yields another improvement.
Most queries run in sublinear time and a few ($<20\%$) run in linear time,
because of dependencies on input source-list and result-set size.
We observed that for this logical database 
partitioning
schema the limiting cadence the 
pipeline achieved with processing IPHAS data is 25\,seconds.
\end{abstract}


\end{frontmatter}


\section{Introduction}
\label{sec:intro}

High-cadence astronomy is a relatively new field in observational astronomy.
Advances in hardware and software technology have made it possible to
stream large volumes of observational data over fast links to 
clusters of computers that, in general, process the data 
in one or more automated pipelines for scientific analysis.
The time available to do real-time analysis
is limited by the cadence of the instrument.
Therefore, additional and complementary scientific data analyses 
are forced to shift to non-real time environments.
Here, all data accumulates over time and 
the growth may vary in the range of 0.1--100~PB/yr 
\citep{Becla}. 
These volumes clearly challenge many aspects of contemporary 
data management systems, which is also recognised by
\citet{Ivezic}.

Several instruments have shown impressive demonstrations of charting 
the sky down to 
a timescale of seconds, 
e.g., 
the international LOFAR telescope \citep{vHW}, 
the Murchison Wide-field Array \citep[MWA;][]{Tingay},
the Australian Square Kilometre Array Path\-finder \citep[ASKAP;][]{Murphy}.
High-cadence observations in image-domain astronomy, 
where sky regions are revisited many times 
in relatively short periods, 
produce overwhelmingly large amounts of data.
Optical and radio telescopes planned for the next decade will
generate even larger continuous data streams,
e.g., 
the Large Synoptic Survey Telescope \citep[LSST;][]{Lazio, Juric}, 
the additional Gound-based Wide Angle optical Camera \citep[GWAC;][]{SVOMGWAC}
of the Space-based multiband astronomical Variable Objects Monitor (SVOM),
BlackGEM \citep{Bloemen},
the Square Kilometre Array \citep[SKA;][]{Broekema}.

Although high-cadence instruments are 
specifically designed to carry out their own unique science, 
they share similar observational strategies.
The main ones being: high-speed, wide- or all-sky surveys, 
searching for transient and variable sources on a variety of time scales
and gradually archiving full-source light-curve catalogues.
In this respect, the archive is considered the new Big Data laboratory,
equipped for making scientific discoveries in complex structured data.
However, such discoveries are only possible when 
the infrastructure and software tools allow 
continuous and simultaneous data mining,
statistical 
modeling, machine learning and ad-hoc querying.

The optical Sloan Digital Sky Survey 
\citep[SDSS;][]{York,Alam}
was the first instrument to seriously integrate a database system 
into its survey design.
It uses a database-centric computing approach for 
their large-scale scientific datasets.
SDSS data are cumulatively released to the 
public in roughly annual cycles.
In this respect, SDSS is a low-cadence instrument since the
yearly updates of the full-source catalogue makes 
the database essentially static.

On account of Gray's law to ship computations to the data
instead of data to the computations \citep{SzBl}
many algorithms are designed to run \emph{inside} the database engine.
Another design rule includes knowledge of the 50 most frequent and
intensive queries. 
Since astronomical pipelines process the data in a structured way, 
this allows one to optimise execution plans for known queries.

In the radio regime, the automated Transients Pipeline (TraP) of the international  
LOFAR telescope adopted many database techniques 
from SDSS \citep{Swinbank}. 
The TraP applies source finding and fitting 
to calibrated radio images after which, 
per image, all image and source properties, 
i.e.\ the \emph{source list},
are handed over to the data\-base.
Note that the images themselves are not stored in the database.
The loop of tasks of the TraP database consists in total of about 
50
queries which 
can be divided into four successive steps, all executed in bulk mode: 
\begin{enumerate}
\item{load source list}
\item{cross-match source list with catalogue of known sources}
\item{update catalogue: maintain up-to-date statistical sky model}
\item{find/identify transient and variable sources or 
other significant deviations from the sky model}
\end{enumerate}

Typical source lists for LOFAR survey-mode observations do not 
exceed $10^3$ entries, whereas averages are less than $10^2$ 
for cadence modes as high as 10 seconds
\citep{Swinbank}.
The total number of unique sources in the LOFAR radio catalogue
is of the order $10^6$.
Long-term monitoring of the database tasks is essential
to predict pipeline performance 
and understand the instrument as a whole.
Queries with poor scaling (e.g.\ exponential) 
will eventually jam the processing. 
Significant increases of cadence and/or source density
determine the critical limits of the system and 
permitted types of observations.
\citet{Swinbank} show that the TraP run times increase linearly
with input size within the LOFAR observation constraints.

Source lists produced by optical instruments are in general much larger,
primarily due to the intrinsic higher resolution in combination with
the 
increased
sensitivity.
Also the catalogues that represent the optical sky models
hold orders of magnitude more sources than their radio counterparts.
Therefore, one avoids naive implementations of the TraP 
for optical instruments, 
since the extrapolation of the source counts into the 
optical spectrum will most probably break linear performance 
or even in a best-case long-term linear performance scenario,
the processing time will pass the cadence time
at some point.

The planned wide-field optical telescope array BlackGEM 
is dedicated to 
measure optical emission from pairs of merging neutron stars and black holes
\citep{Bloemen}.
BlackGEM will start with 3 telescopes,
all of which will be located 
at ESO La Silla, Chile. 
MeerLICHT,\footnote{Check the current status at \url{www.meerlicht.org}}
a single BlackGEM telescope acting as a prototype,
is
coupled to the MeerKAT radio array 
\citep[a precursor to SKA;][]{Brederode}
to operate in concert and allowing to study the optical--radio sky simultaneously
as a true multi-wavelength instrument.
Table~\ref{tab:BGMLchar} shows the characteristics 
of a single BlackGEM telescope.
From a database--pipeline perspective, the most influential properties
are the source data size, 
the source density
and the integration time,
where the latter determines the cadence.

\begin{table}[!t]
\begin{tabular}{ll||c|c}
\hline
\multicolumn{2}{l||}{Mirror diameter} & \multicolumn{2}{c}{65\,cm}  \\
\multicolumn{2}{l||}{FoV} & \multicolumn{2}{c}{2.7\,deg$^2$}  \\
\multicolumn{2}{l||}{CCD size} & \multicolumn{2}{c}{10,536\,$\times$\,10,536} \\ 
\multicolumn{2}{l||}{Resolution} & \multicolumn{2}{c}{$0.57''$/px}  \\
\multicolumn{2}{l||}{bits per px} & \multicolumn{2}{c}{16}  \\
\multicolumn{2}{l||}{Image size} & \multicolumn{2}{c}{222\,MB}  \\
\multicolumn{2}{l||}{Calib.~images per night} & \multicolumn{2}{c}{2$\times$(10\,bias + 5$\times$5\,flats) = 70}  \\
\multicolumn{2}{l||}{DB source data size} & \multicolumn{2}{c}{402\,B}  \\
\hline
\multicolumn{2}{l||}{Observation mode} & nominal & fast \\
\hline
\multicolumn{2}{l||}{Integration time} & 5\,min & 1\,min \\
\multicolumn{2}{l||}{Sensitivity} & 23\,mag & 21\,mag \\
\multicolumn{2}{l||}{Science images per night} & 120 & 600 \\
\multicolumn{2}{l||}{Data rate per night} & 42\,GB & 148\,GB \\
\hline
\end{tabular}
\caption{Characteristics of the MeerLICHT telescope,
a single BlackGEM prototype telescope.
DB source data size is the storage size that all properties
of single source would need when stored in a database.
We assume observation nights of 10\,hrs.
}
\label{tab:BGMLchar}
\end{table}

In this paper we 
use the TraP-like queries as a baseline and
investigate its scalability 
to the MeerLICHT environment.
We need to know to what extent the processing of 
images with 
source densities of 500,000 deg$^{-2}$
or source lists with hundreds of thousands of sources 
is still feasible.

The paper is outlined as follows.
Section~\ref{sec:db} gives the rationale 
behind the choice of a column-oriented relational database management system
(RDBMS).
Section~\ref{sec:perf} describes the 
experimental set up of the performance tests,
the data 
and the TraP queries that were adjusted and optimised.
The results are presented in Section~\ref{sec:res} 
and concluded in 
Section~\ref{sec:concl}.
Although all source codes are publicly available,
the Appendices show the relevant query snippets for readability.
\\

\section{Rationale for MonetDB, a column-oriented 
relational database management system
}
\label{sec:db}

Non-relational databases, e.g., key-value and NoSQL stores, lack 
solid support for many data manipulations that are required 
for this type of astronomical application.
We cannot afford redundant or duplicate storage,
and therefore have to distribute the data over
a minimal set of related tables.
The absence of fast join and cross-match functionality,
schema free storage formats, no transactional support 
and own unique query languages
makes non-relational databases
hard to perform on such applications. 
Furthermore, 
the bulk processing requires fast data aggregation, 
ordering and indexing 
to avoid large table scans. 
All these functionalities are well-known and implemented 
in relational database systems.

Relational database storage models 
follow either the 
row-oriented or column-oriented principles.
The row\hyp{}oriented storage model partitions tabular data horizontally.
Such record layouts consist of rows that store all their columns contiguously,
with the adverse effect that a data block contains multiple data types.
Queries that touch only a few columns of a large table or
joins of tables,
waste both bandwidth and memory space in this model,
because the blocks that the CPU reads and buffers are contaminated
with all the other, unwanted, columns.
\citet{Abadi08} discuss more differences between row and column stores.

The columnar model splits tabular data vertically. 
Every column is represented by an array of single data type.
These kinds of array can be manipulated easily and scanned quickly 
when sorted.
Its variables of fixed size are densily stored, possibly in compressed format, on disk.
The uniformity of the array fits well to the block-oriented nature 
of memory transfers and CPU caches
and exposes good spatial locality 
and high cache hit ratios
when queries execute large scans over a subset of columns.
More in-depth details of column-oriented databases can be found 
in \citet{Abadi12}.

More research and developments in the field of column-oriented databases 
led to the implementation of many related techniques in the open-source
main-memory relational database management system MonetDB \citep{Boncz}.
MonetDB's architecture is geared toward read-optimised data\hyp{}intensive scientific applications
\citep{Marcin}.
It is compliant with the SQL-2008 standard
and has language bindings for \texttt{C}, \texttt{Java}, \texttt{Python},
\texttt{R}
and \texttt{Java\-Script/Node.js}.
The ease of extending its functionality with 
user-defined functions (UDFs) written in \texttt{SQL}, 
\texttt{C}, \texttt{R} and \texttt{Python} are other serious strengths. 
MonetDB follows a strict columnar design and takes into 
account the underlying computer architecture
\citep{BMK99, Heman}.
The fundamental removal of the expression interpreter 
fully eliminates parsing expensive code,
verifying record layouts and checking data types.
Hard-coded semantics makes MonetDB's algebra simple yet
efficient, because all operators work on simple arrays
allowing the compiler to generate CPU-friendly instructions.
Column-at-a-time processing reduces 
the number of function calls
and data and control dependencies,
which in turn improve the algorithms and the CPU cache performance,
as opposed to tuple-at-a-time iterators.
Look-up and range queries benefit from 
hash-indexes and secondary \emph{imprints} indexes, resp.,
that are automatically built for touched columns
\citep{Lefteris}.
A query optimiser generates and analyses alternative
query plans and executes the plan that minimizes the query cost.
\citet{Stefan} developed cost models that 
estimate the query execution time based on I/O and average CPU costs.
\citet{Ivanova13} extended MonetDB's code base
to support SQL management of external data (SQL/MED).
Loading binary columnar catalogue \texttt{FITS} files 
in this way
is orders of magnitude faster 
than using classical 
statements,
since the in-memory binary data exactly matches MonetDB's storage model.

Until now, the largest database archive for a single telescope is 
the SDSS SkyServer 
tuned to the commercial closed-source row store Microsoft SQL Server.
Full-sized data releases were successfully ported into MonetDB.
\citet{Ivanova} demonstrated that MonetDB is capable of loading 
and querying the SDSS SkyServer data.
Initial performance evaluations on a smaller subset indicated that
85\% of the most executed queries
run faster in MonetDB, while the remaining are of competitive speed.
Most of the time scientific queries only touch a few columns,
whereas the tables generally have many hundreds of columns.
This is a strong call for using column-oriented databases 

In all our experiments we use the Structured Query Language (SQL)
to interact with the data. 
SQL queries access data directly and return aggregated or full result sets,
in contrast to scripts that retrieve data sets and 
process tuples iteratively. 
(Note that in this context a query is a generic term for any kind of 
instruction set(s) that run on database data.)
Query response times are critical in high-cadence pipeline applications.
Because it is more efficient to process data in bulk mode, 
the fact that data access is fastest close to the CPU
and that astronomical queries very often work on columns or ranges thereof,
the choice for a main-memory column-oriented database 
in astronomical pipelines is obvious.

During database kernel and pipeline query development we
ran comparison tests with main alternative open source RDBMSs regularly.
However, reports of these results are beyond the scope of this paper.

\section{Baseline and alternative high-cadence pipeline benchmarks}
\label{sec:perf}

\subsection{BlackGEM pipeline architectures}
\label{sec:pipearch}

BlackGEM will produce about 1\,TB of data per night
of which 90\% are raw and calibrated images and 10\% 
extracted information for the full-source database.
Raw data are calibrated and imaged after which
the source-extraction output product is 
a list of all detected sources and their properties
in binary catalog FITS format.

A primary image-differencing pipeline runs in real time 
to detect transient and variable events in the data stream
of calibrated images. 
For differencing it uses reference images and
parts of the source list for PSF fitting.
The output product is a binary catalog FITS file of  
all transient sources, which will be stored in a separate
relatively small transient-source database,
which will not be discussed any further here.

The full-source binary catalog FITS files
serve as input for a secondary
pipeline that runs in an offline mode
and consistently stores all sources into the full-source database for 
scientific analysis.
Delays in this mode are acceptable up to the point where
the overall cadence is not being met anymore.
This paper concentrates on the secondary pipeline
and its full-source database.
The high cadence and source densities of BlackGEM and MeerLICHT
force us to carefully monitor the long-term run-time performance 
of the pipeline queries,
especially since they run in a dynamically growing database.
The pipeline processes the data in a number of steps described in \S~\ref{sec:intro}.
After the sources have been loaded into the database (step 1),
read, write and delete queries take care of the source association
and sky model maintenance procedures
(steps 2 and 3).

It needs to be noted here that 
transactions in MonetDB follow the optimistic concurrency control scheme.
The lack of a locking scheme implies that queries modifying the data 
(of which we have many) are serialized at the application level
and run in single-threaded mode.
Explicitly programmed multiple threads with own database connections
can run in parallel easily if the query's write operations 
are executed on independent tables, however,
we did not code that in our modules.
On the other hand, queries that only read data 
run in parallel multi-threaded mode implicitly.

\subsection{IPHAS data}
\label{sec:iphas}

Before BlackGEM is operational
we process real binary catalogue \texttt{FITS} files 
from the Isaac Newton Telescope (INT) Photometric 
H$\alpha$ Survey of the Northern Galactic Plane 
\citep[IPHAS;][]{Barentsen}.
The $10\sigma$ limit is at magnitude 20 and the field of view is 0.29~deg$^2$.
IPHAS has about 200 observations per night, 
where single source lists range between 1,000 and 130,000 entries, 
with peaks up to 170,000.

\subsection{The SciLens cluster}

We run our experimentation queries on multiple nodes of the 
SciLens\footnote{\url{http://www.scilens.org}}\footnote{For 
the current status and overview of the configuration, see
\url{https://www.monetdb.org/wiki/Scilens-configuration-standard}}
cluster located at CWI.
Our queries ran on single cluster nodes, 
to which the input files were transfered.
Table~\ref{tab:SciLconfig} gives an excerpt of the specifications
of the nodes that we used.

\begin{table*}[!t]
\centering
\begin{tabular}{ll||c|c|c|c}
\hline
\multicolumn{2}{c||}{}    & \textbf{diamonds} & \textbf{stones} & \textbf{bricks} & \textbf{rocks} \\
\hline
\hline
\textbf{CPU}   & Architecture   & \multicolumn{4}{c}{x86\_64}  \\
\cline{3-6}
 & CPU(s)   & 96 & \multicolumn{2}{c|}{32} & 8 \\
\cline{3-6}
 & Threads per core   & \multicolumn{4}{c}{2}   \\
\cline{3-6}
 & Cores per socket   & 12 & \multicolumn{2}{c|}{8} & 4 \\
\cline{3-6}
 & Socket(s)   & 4 & \multicolumn{2}{c|}{2} & 1 \\
\cline{3-6}
 & Clockspeed & $2.4-2.9$\,GHz & $2.6-3.4$\,GHz & $2.0-2.8$\,GHz & $3.4-3.8$\,GHz\\
\hline
\textbf{RAM} & Size   & 1024\,GB & \multicolumn{2}{c|}{256\,GB} & 16\,GB\\
\hline
\textbf{SSD} & Drives   & \multicolumn{2}{c|}{} & $8\times128$\,GB & \\
\cline{3-6}
             & \texttt{/ssd}   & \multicolumn{2}{c|}{} & 8x HW RAID0    & \\
\hline
\textbf{HDD} & Disks   & $4\times2$\,TB & $3\times3$\,TB & $4\times2$\,TB & $1\times2$\,TB\\
\cline{3-6}
             & \texttt{/scratch} & 7.2\,TB       & \multicolumn{2}{c|}{5.4\,TB} & 1.8\,TB \\
             &                   & ($4\times$\ HW RAID0) & \multicolumn{2}{c|}{(3$\times$\ SW RAID0)} & \\
\cline{3-6}
             & \texttt{/data}   &   & 3$\times$0.9\,TB & 1.8\,TB & \\
\hline
\textbf{Network} & Ethernet   & $2\times10$\,Gb/s & \multicolumn{3}{c}{1\,Gb/s} \\
\cline{3-6}
                 & Infiniband & $4\times40$\,Gb/s & \multicolumn{3}{c}{40\,Gb/s} \\
\hline
\textbf{Software} & OS & \multicolumn{4}{c}{Linux, Fedora 24 4.7.3-200.fc24}    \\
\cline{3-6}
                  & MonetDB & \multicolumn{4}{c}{Jun2016 SP1}    \\
\cline{3-6}
                  & Python & \multicolumn{4}{c}{2.7.12}    \\
\hline
\end{tabular}
\caption[]{Configuration of SciLens cluster nodes}
\label{tab:SciLconfig}
\end{table*}


\subsection{Baseline pipeline and query monitoring}
\label{sec:monitor}

We 
used
the May 2007 binary catalogue \texttt{FITS} files as 
input for the prototype pipeline.
The series consists of 1893 files, where each \texttt{FITS} file has 
four extensions
due to the four CCDs of the telescope.
The \texttt{Python} prototype pipeline script is based on the TraP \citep{Swinbank}
and was modified for the load and 
cross-match stages (steps 1 and 2)
before it could serve as the baseline pipeline.

The most significant change in the data loading of step 1 was the replacement
of the classical SQL insert statements with 
SQL/MED queries \citep{Ivanova13}. 
This technique \emph{attaches} datasets to the database that can be queried
before they are actually loaded. 
Larger files, i.e., longer source lists, can be handled, because
the data do not need to be parsed.
In step 2, the variable conical search radius that is being used 
for cross-matching radio
sources was set as a constant parameter because of 
the more stable quality of optical images
(see \S~\ref{sec:cm} below).

All individual query run times are written to log files
for easy plotting
and identification of CPU and I/O intensive queries.
Sets of queries that belong to a step as described in \S~\ref{sec:intro}
and \S~\ref{sec:pipearch}
are grouped together and prefixed \texttt{I} or \texttt{A}
for loading/inserting or cross-matching/associating, resp. 
The modified 
TraP, as described in the previous paragraph,
is considered as the baseline 
pipeline
to which the query optimisations will be compared.

All source code presented in this paper is publicly and freely 
available for download and usage at the CWI  
\texttt{git}-repository
site.\footnote{\url{https://scm.cwi.nl/DA/blackgem-code}. The git version
used for this paper was \texttt{3ad947789a19}.}
It also provides instructions to get started.

\subsection{Cross matching and baseline module \texttt{M0}}
\label{sec:cm}

We adopted the set of TraP source association queries,
including the cross-match query,
and modified it slightly 
for the processing of optical data.
This then is our baseline module \texttt{M0}.
The source association queries are the most intense and sensitive database operations,
where a source list 
is cross-matched with the internal catalogue of known sources.
At the core of this query 
is the positional source look-up by a conical search.
From SDSS SkyServer log analyses 
\citet{Ivanova} found 
that these conical searches
were part of the most frequently executed queries.
Optimising this will minimize CPU time and
compensate processing times for other queries.

The cross-matching method applied here is based on the 
TraP source association 
described in \S~4 of \citet{Swinbank}.
In that paper it was also shown that the association algorithms scale 
reasonbly linear with the number of sources.
Source lists in their performance tests, however,  
do not exceed 1200 entries, which is acceptable in 
the radio domain of LOFAR, but not
in the optical regimes of MeerLICHT and BlackGEM.
The TraP is untested for source densities 
that are orders of magnitude larger,
making it plausible that we cannot 
simply extrapolate the TraP test results
to the optical domain.

The resolution and positional uncertainties in the optical IPHAS images are more stable 
compared to the LOFAR case, where the image resolutions 
fluctuate due to more complex radio-specific calibrations and antenna dependencies.
Therefore, the TraP implements a variable search radius to cross-match sources,
but because the optical image quality is more stable 
we can replace it in the query algorithm
by a simpler conical search radius of constant value 
that is determined by the telescope's resolution element.

The cross-match baseline query is written out in 
\ref{app:q0}.
It joins the sources from the latest appended source list
available in the \texttt{extractedsource} table
with the known catalogued sources stored in \texttt{runningcat\-a\-log}.
If the distance between a found source pair is less than 
the search radius the pair is considered as a genuine association
and both \texttt{ID}s and distance are returned.
In fact, however, more properties are returned, but for 
illustrative purposes they are omitted here.
It must be noted here that the cross-match query result set 
does not contain only unique catalog source--extracted source pairs.
Multiple types of association pairs are possible, falling into
the categories of \emph{no association}, \emph{one-to-one},
\emph{many-to-one}, \emph{one-to-many} and \emph{many-to-many}.
The association module takes care of further processing these topologies,
which are described in detail by \citet{Swinbank}.

The on-sky distance is only calculated for 
catalogue counterparts that lie within 
the boxes centred at the source positions from the list. 
The widths of the boxes are determined by the fixed search radius.
Both tables have an integer \texttt{zone} column of 8-bit data type,
that specifies the declination strip in which the source lies.
The box height gives which neighboring declination strips needs to be searched.
The on-sky search radius is constant, but expressed in RA
it varies depending on declination.
Therefore, the RA-box width increases when moving towards the celestial poles.
The user-defined \texttt{alpha()} function  
determines the rate of inflation as given by 
\citet{Gray}.

The distance in radians on the sky, $\vartheta$, between a measured source position,
$\mathbf{x}$,
and its candidate counterpart in the running catalogue,
$\mathbf{m}$,
is given by the dot product $\mathbf{x}^T\mathbf{m}=\cos\vartheta$,
where $\mathbf{x}$ and $\mathbf{m}$ are unit vectors.
However, when dealing with small angles the alternative 
of using the sine function gives computationally more accurate results.
Therefore we use the arc-angle distance between $\mathbf{x}$ and $\mathbf{m}$
to determine $\vartheta$
\begin{equation}
\sin\tfrac{1}{2}\vartheta = \tfrac{1}{2} ||\mathbf{x} - \mathbf{m}||,
\end{equation}

The above mentioned box size and distance criteria
are declared in the cross-match baseline SQL statement 
(see \ref{app:q0}).
Naive SQL execution plans in the case of large source lists 
may downgrade the performance to unacceptable levels. 
Therefore, we have to carefully design several alternative cross-match queries 
that implement different search techniques 
and corresponding query optimisations in order to evaluate their
performance under different circumstances.

In the next paragraph we will elaborate on alternative queries, 
in a database schema where we 
partition and sort the largest tables by declination.

\begin{figure*}[!t]
\centering
\includegraphics[width=0.35\hsize,valign=t]{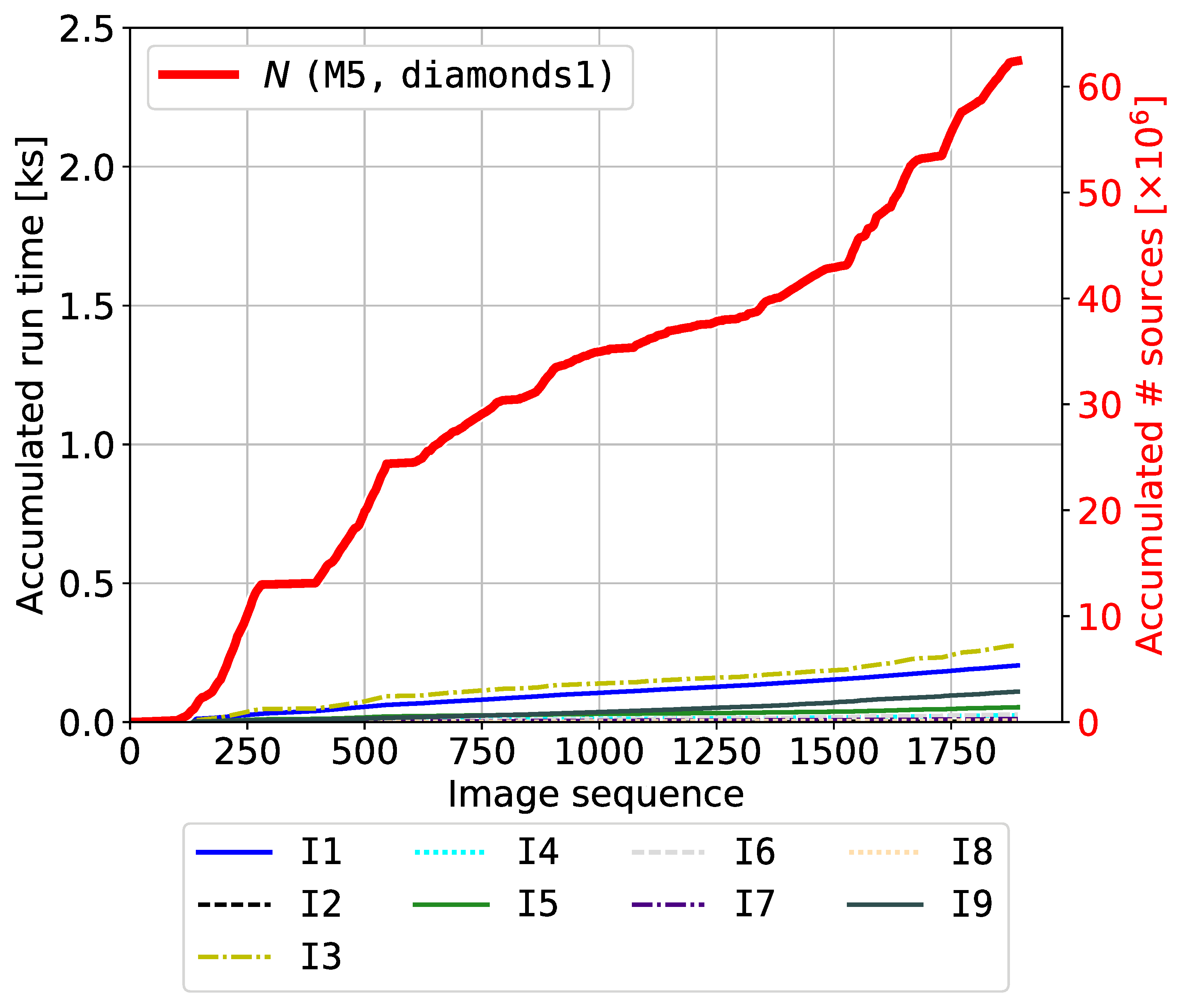}%
\includegraphics[width=0.6\hsize,valign=t]{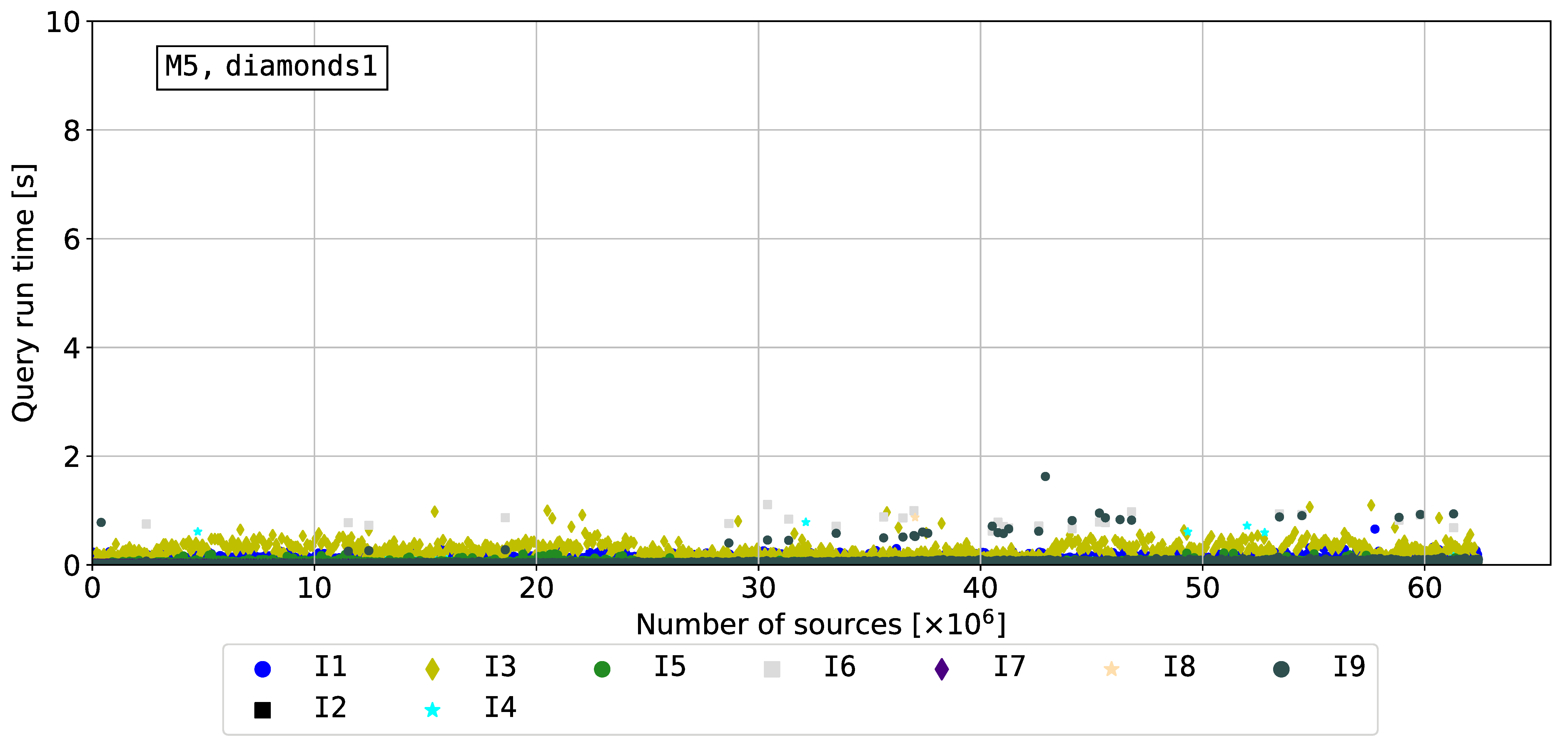}
\caption{Accumulated (left-column graphs) and 
instantaneous (right-column graphs) run times of the loading queries 
in the alternative module \texttt{M5} 
on the diamonds node in the SciLens cluster
with respect to the image sequence number and 
the number of sources appended to the database, resp.
(See Table~\ref{tab:SciLconfig} for the node specifications.)
The thick (red) line in the left-column graphs shows 
the growth of the appended number of sources,
which is a measure of the database size.
\texttt{I15--18} is the subset of queries that is responsible 
for the database schema clean up.}
\label{fig:i_Q0}
\end{figure*}

\subsubsection{Alternative module \texttt{M5}}
\label{sec:descr_M5}

Data partitioning is a well-known concept in database 
designs to control load balancing and performance. 
Tables are divided into independent smaller (sub)tables,
which can be accessed at a finer level of granularity.
This functionality allows a table to be defined as a union of its partitions.
One then can query the parent table as if it is an autonomous 
table.\footnote{%
More on the internals can be found at 
\url{https://www.monetdb.org/Documentation/Cookbooks/SQLrecipes/DataPartitioning}.}
On the other hand, for finer control and better data locality one can query 
individual partitions, which are tables,
to avoid access to partitions that are outside the query.
In module \texttt{M5} we investigate the performance of the cross-matching pipeline
where the largest tables are partitioned
horizontally into declination zones of one degree.

We force the 
partition tables to be sorted according to declination,
which uses the system cores in parallel.
This means that when table rows need to be updated or appended, 
the partition table is rewritten. 
This is acceptable as most of the table chuncks are sorted 
according to their original format.
Moreover, cross-matching will be 
fast and can be done in memory,
since the tables are relatively small and in sorted order.
The partitioning scheme is also prepared to withstand database growth,
since the partition sizes do not exceed the expected number of 
sources.
The relevant \texttt{M5} SQL code is shown in \ref{app:M5}.

\section{Presentation of experimental results}
\label{sec:res}

The baseline (\texttt{M0}, \S~\ref{sec:cm}) and alternative (\texttt{M5}, \S~\ref{sec:descr_M5}) modules 
both have the same set of SQL/MED loading queries,
but a different set of cross-matching queries.
On the same type of node 
the same loading queries compete for memory and CPU
with different sets of cross-matching queries from the respective modules.
Every cross-matching module has its specific methods of memory allocation
and data storage, which produce unequal loading behaviour.
Nevertheless, the variations between modules are small and therefore 
\S~\ref{sec:results_loading} only presents
the loading query run times of the
alternative module \texttt{M5}.
In \S~\ref{sec:results_cm} we present the details of different 
cross-matching modules
and in \S~\ref{sec:results_overall}
the overall performances.

\subsection{Loading part}
\label{sec:results_loading}

The loading part consists of nine queries
that first attach and load the binary \texttt{FITS} files into temporary tables
that are created on the fly (\texttt{I1}--\texttt{I3}),
then copy the \texttt{FITS} header data 
and sources
over into permanent tables (\texttt{I4} and \texttt{I5}, resp.)
and finally clean up the schema (\texttt{I6}--\texttt{I9})
before the next file comes in.
The accumulated run-time performances
of the individual queries 
and the growth of the number of sources in the database
with respect to the \texttt{FITS} image sequence
for the alternative module  
\texttt{M5} on the diamonds 
node are shown 
in the left graph of
Fig.~\ref{fig:i_Q0}.
The thick (red) line in 
this graph
represents 
the number of sources in the database at the moment of query execution and
corresponds to the number of entries in the largest table, 
which is a measure for the database size. 
Although the number of sources increases irregularly,
the individual query run times accumulate linearly over time,
meaning a query runs equally fast at any moment.

The constant-time complexity $\mathcal{O}(1)$ is clearly visible 
in the right 
graph of
Fig.~\ref{fig:i_Q0},
where the individual query run times are plotted 
versus the number of sources in the database.
Although there is some scatter on the nodes that have limited memory
in combination with the slower HDDs (not plotted here), 
these queries run independent of the database size.
It is not possible to make a fair comparison with the loading query set 
of the TraP as presented by \citet{Swinbank}, because they 
do not separate the loading from the cross-matching in their 
performance plots. 
However, the flat performance of the SQL/MED queries for the 
baseline and alternative modules allow the loading 
of source lists that more than 100 times larger.

Queries \texttt{I3}, \texttt{I4} and \texttt{I5}
(spelled out in \ref{app:i3}, \ref{app:i4} and \ref{app:i5}, resp.)
contribute most 
to the total load time.
Query \texttt{I3} is an SQL/MED procedure call.
After the \texttt{FITS} data have been attached
to the database by generating a fully queryable temporary table,
this call really loads the data into the database, 
effectively making the data-vaults table permanent in the 
database schema.
Query \texttt{I4} loads the header data,
common to all sources originating from the same \texttt{FITS} file,
into the \texttt{image} table as a single entry,
whereas \texttt{I5} appends all data from the 
data-vaults table to the permanent
\texttt{extractedsource} table.
Such bulk inserts vary, but can be over $10^5$ entries (see \S~\ref{sec:iphas}).
The insertion of only 
one entry
into the \texttt{image} table
is an expensive operation,
because all fifteen data types per entry have to be parsed.
Comparison to the other append/insert queries (\texttt{I1}, \texttt{I3} and \texttt{I5})
makes the effect more conspicuous,
where the cost of appending $10^5$ sources is low 
since the SQL queries are aware of attaching binary data 
and the data types.
However, all load queries run in constant time,
independent of the database 
size with the ability to scale up even further.

\subsection{Cross-matching part}
\label{sec:results_cm}

After the sources from a \texttt{FITS} file have been inserted 
the cross-matching procedure starts and
runs 15 to 46 queries 
depending whether the baseline or alternative module was chosen.
The source list is cross-matched with counterparts
in the catalog of known objects.
As described in \S~\ref{sec:cm} this results in a candidate list that is
further sifted by subsequent queries to 
resolve the various association types of source--object pairs that turn up.
\citet{Swinbank} discuss the different types and elaborate on
the devised steps, i.e., queries, to append the new source measurements
to existing or new light curves.
The final step is to update the statistical properties of the known 
catalog sources to include the new measurements in the model.
Sources for which no counterparts were found are appended as new
entries in the catalog.

There are many queries involved in this module, but we will focus on
the query that performs the cross-matching.

\subsubsection{The baseline cross-match module \texttt{M0}}
\label{sec:res_M0}

\begin{figure*}[!p]
\centering
\includegraphics[width=0.29\hsize,valign=t]{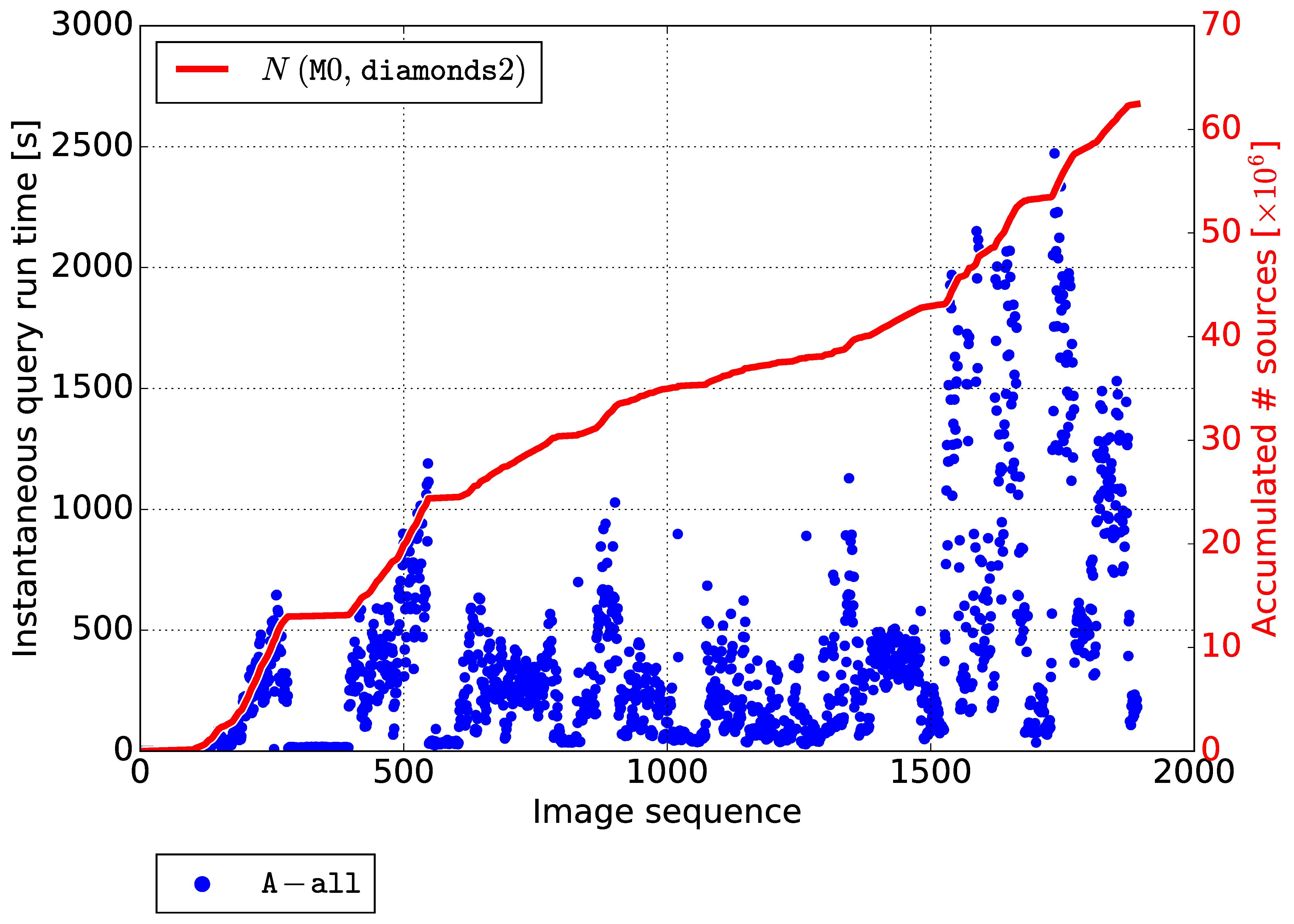}%
\includegraphics[width=0.29\hsize,valign=t]{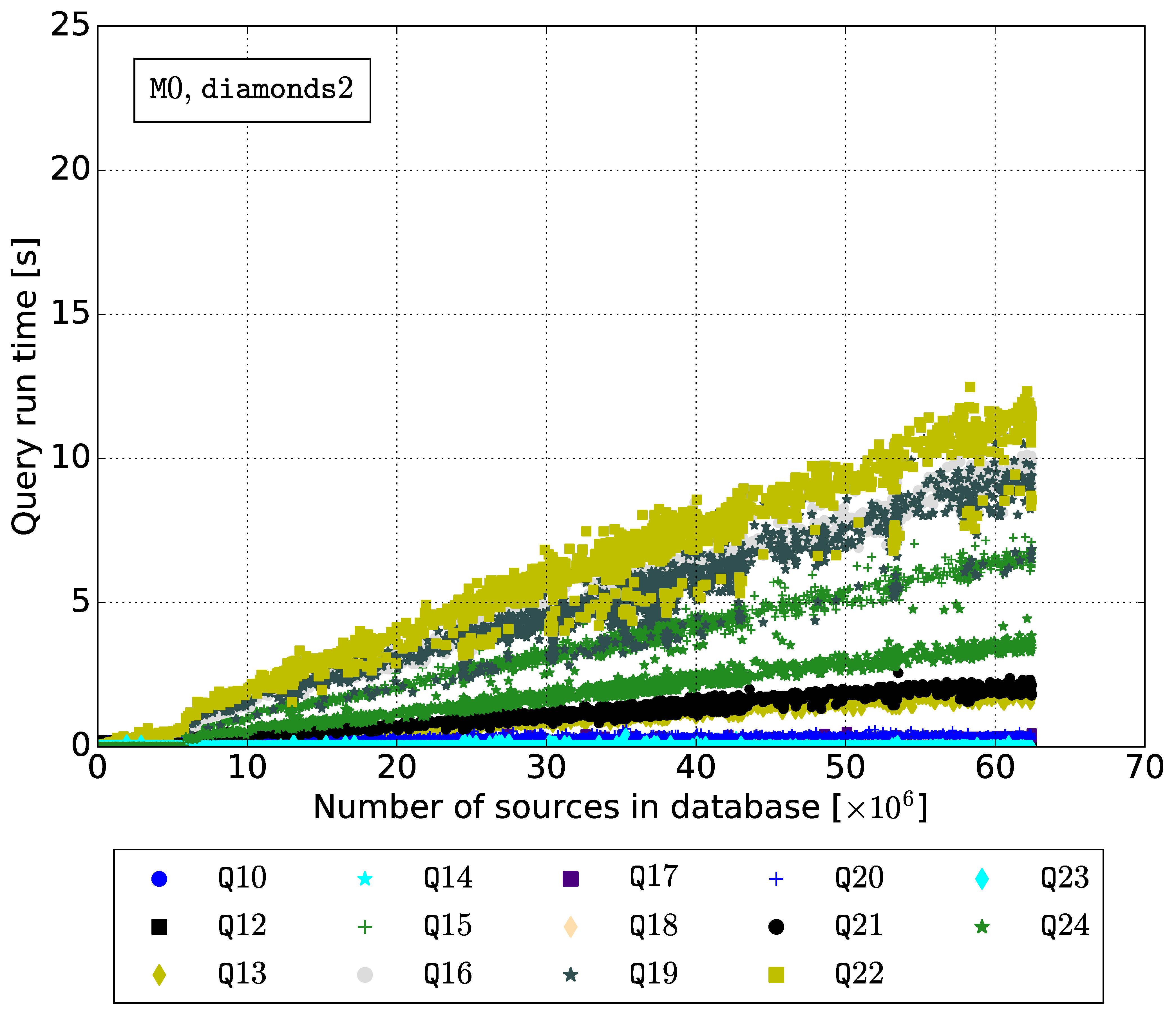}%
\includegraphics[width=0.29\hsize,valign=t]{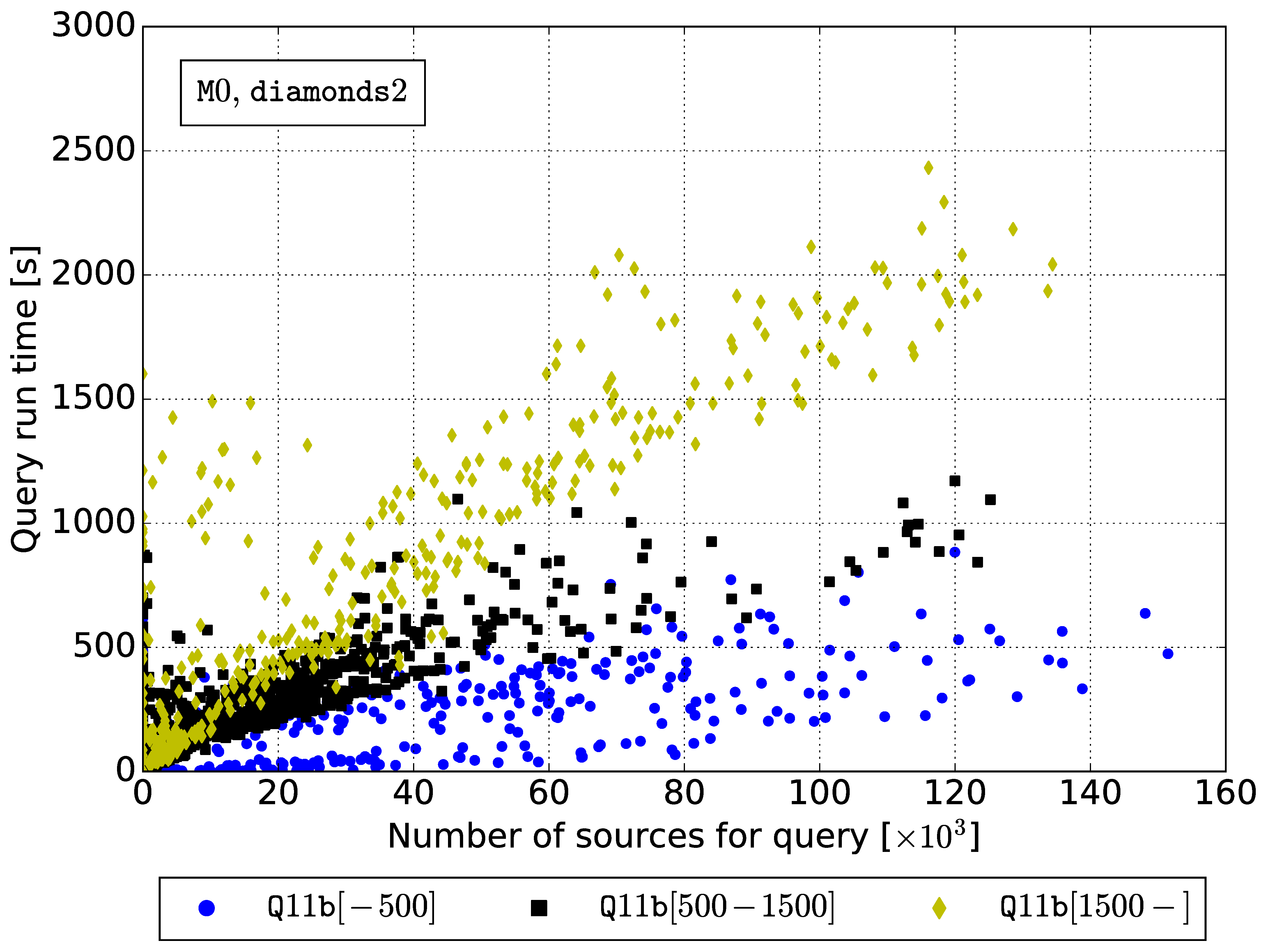}
\includegraphics[width=0.29\hsize,valign=t]{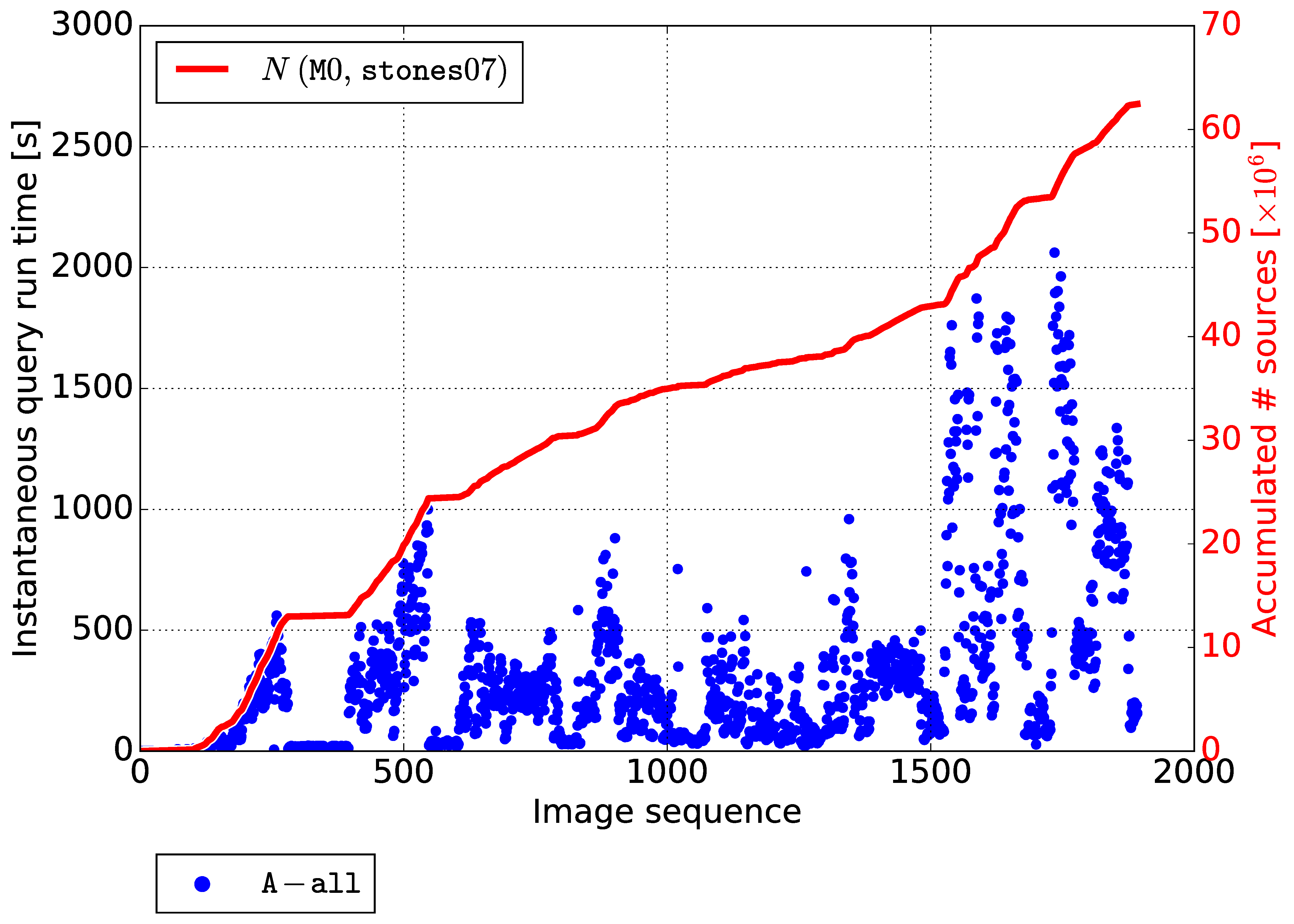}%
\includegraphics[width=0.29\hsize,valign=t]{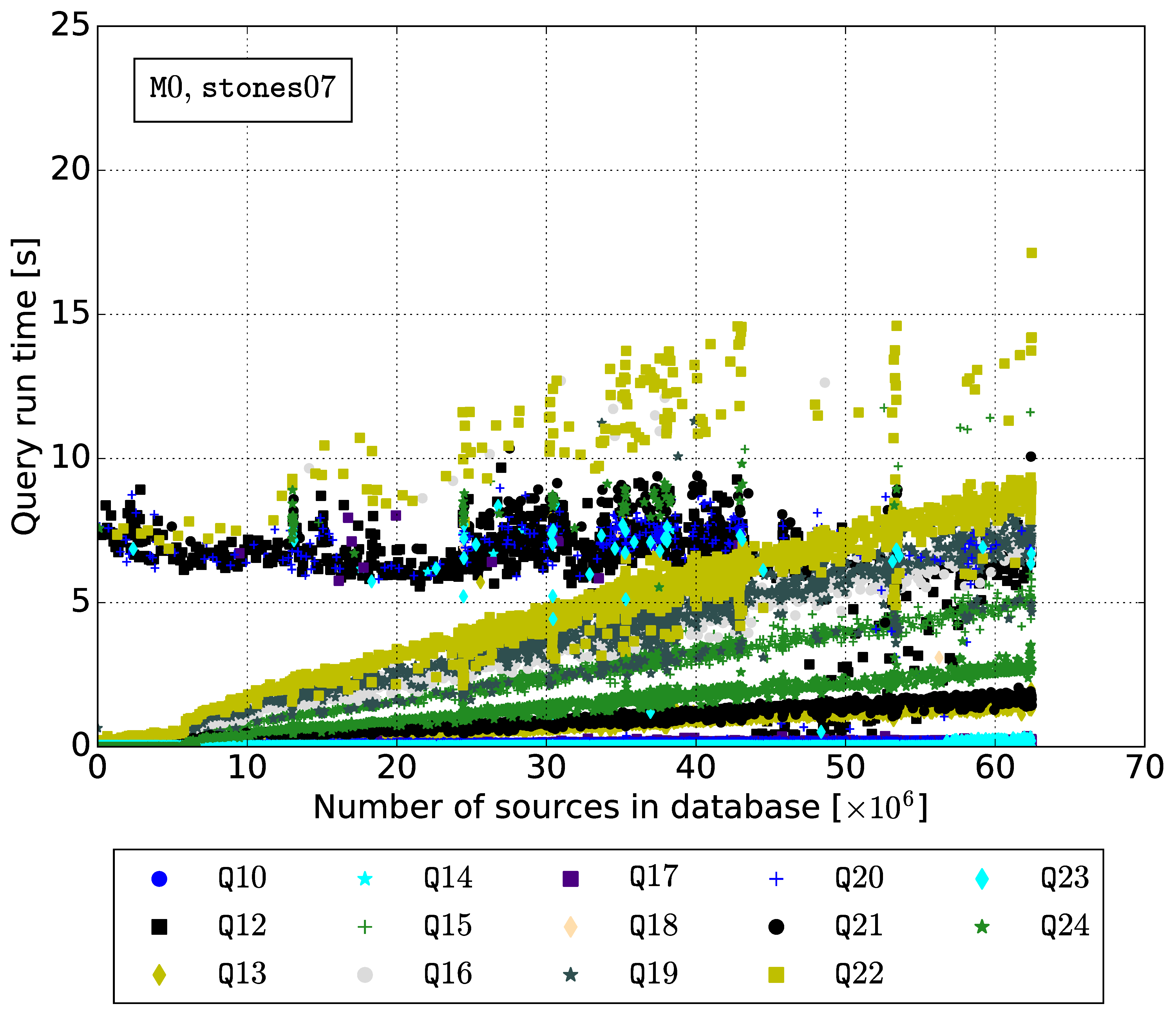}%
\includegraphics[width=0.29\hsize,valign=t]{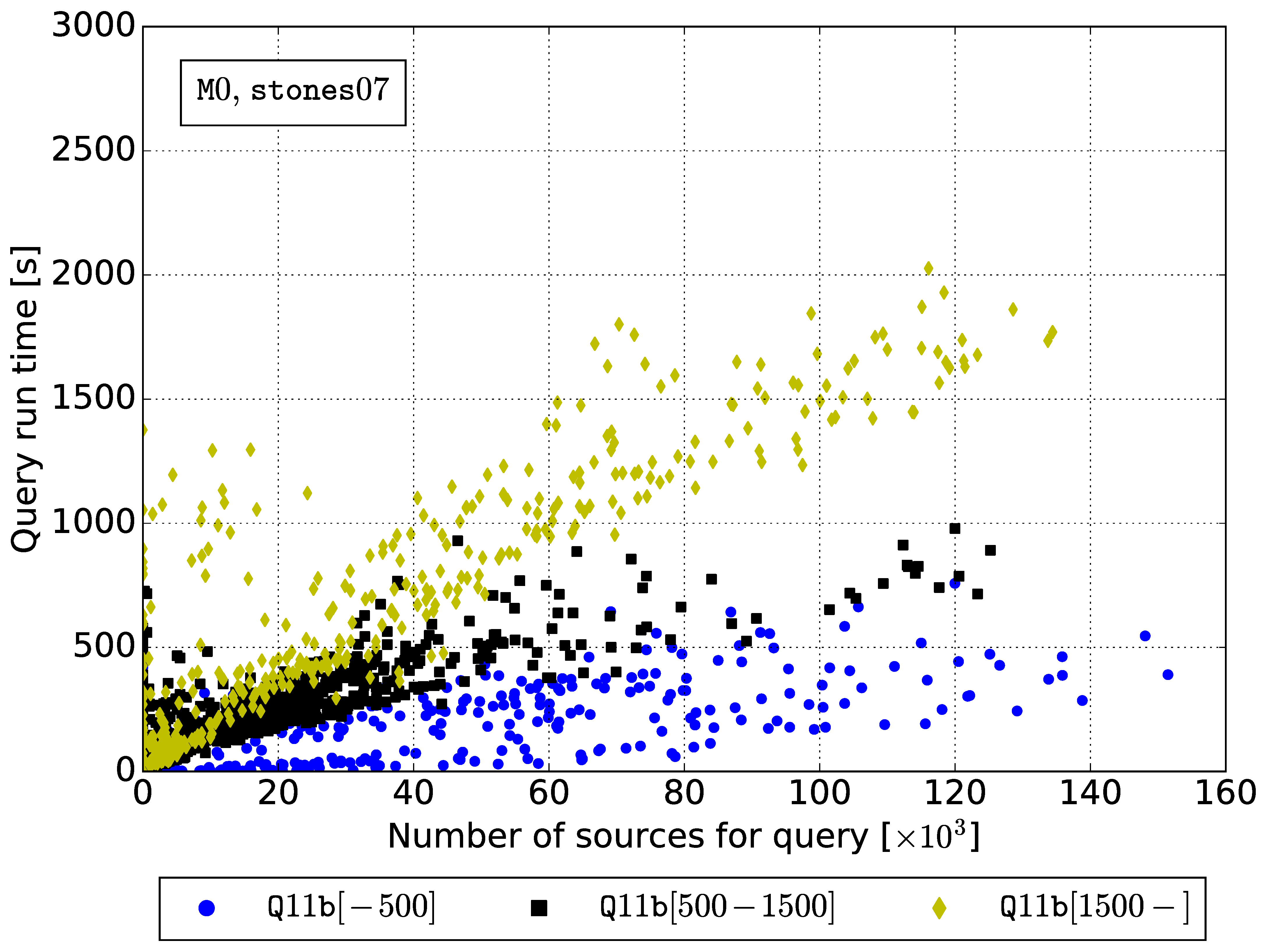}
\includegraphics[width=0.29\hsize,valign=t]{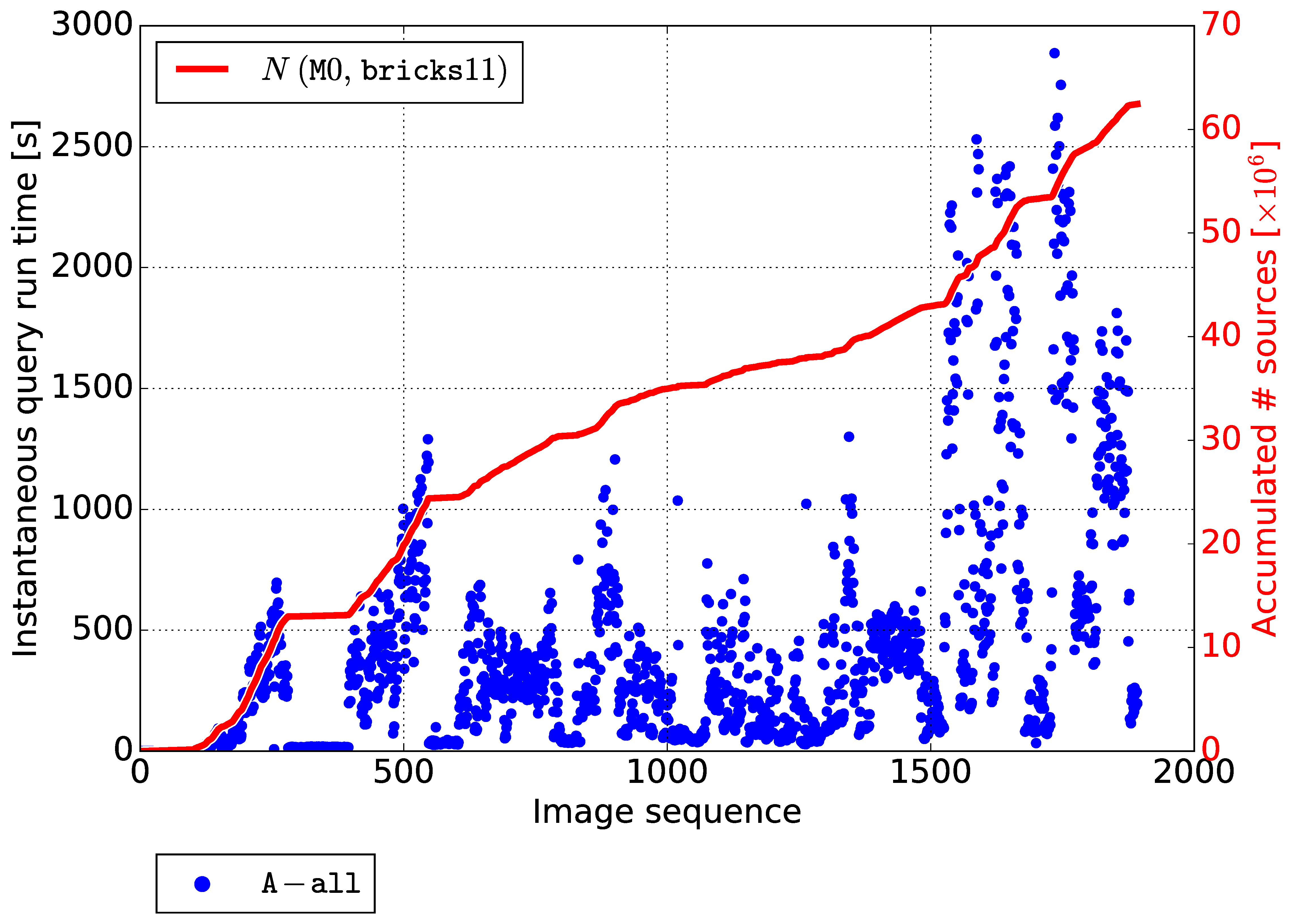}%
\includegraphics[width=0.29\hsize,valign=t]{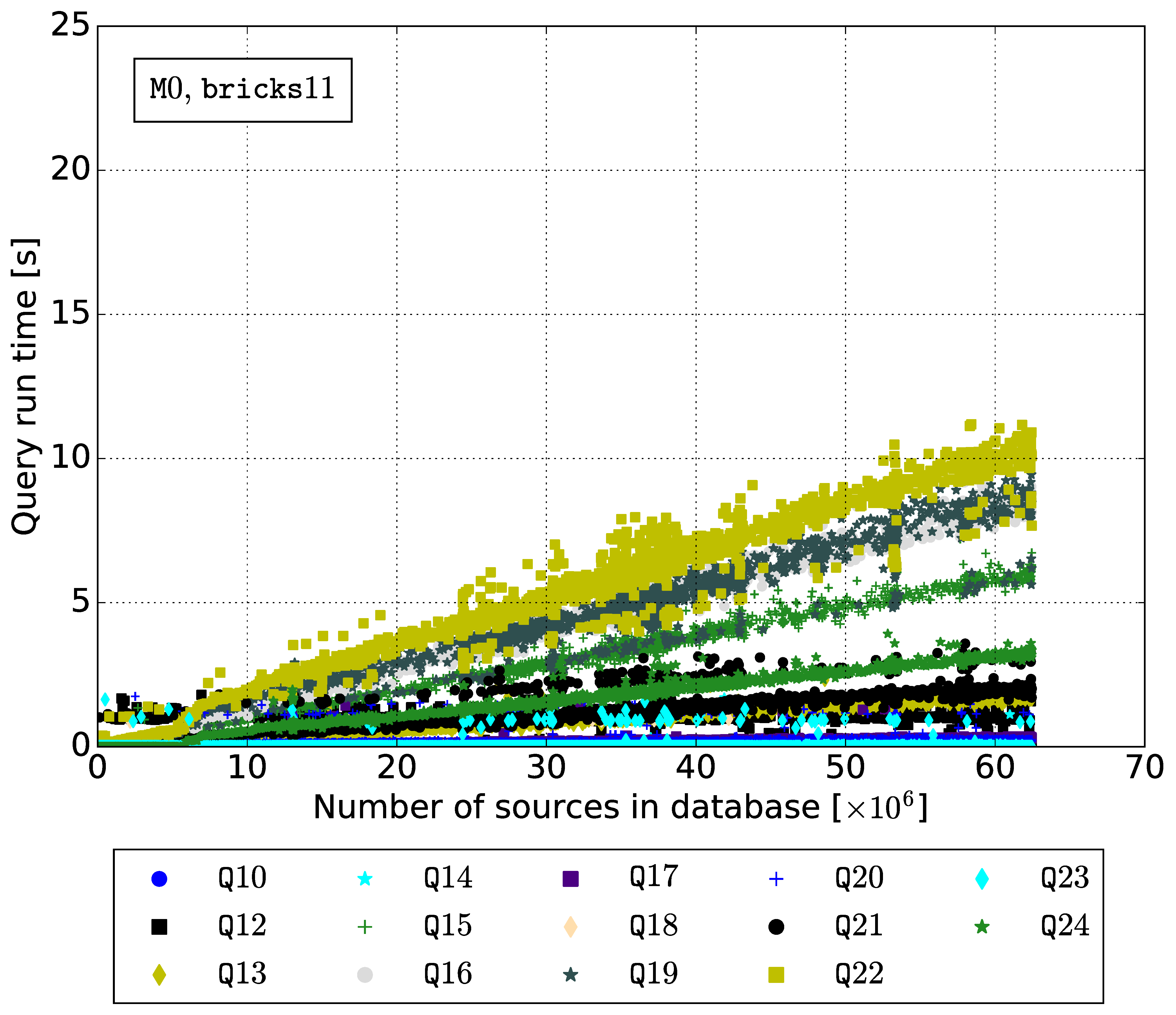}%
\includegraphics[width=0.29\hsize,valign=t]{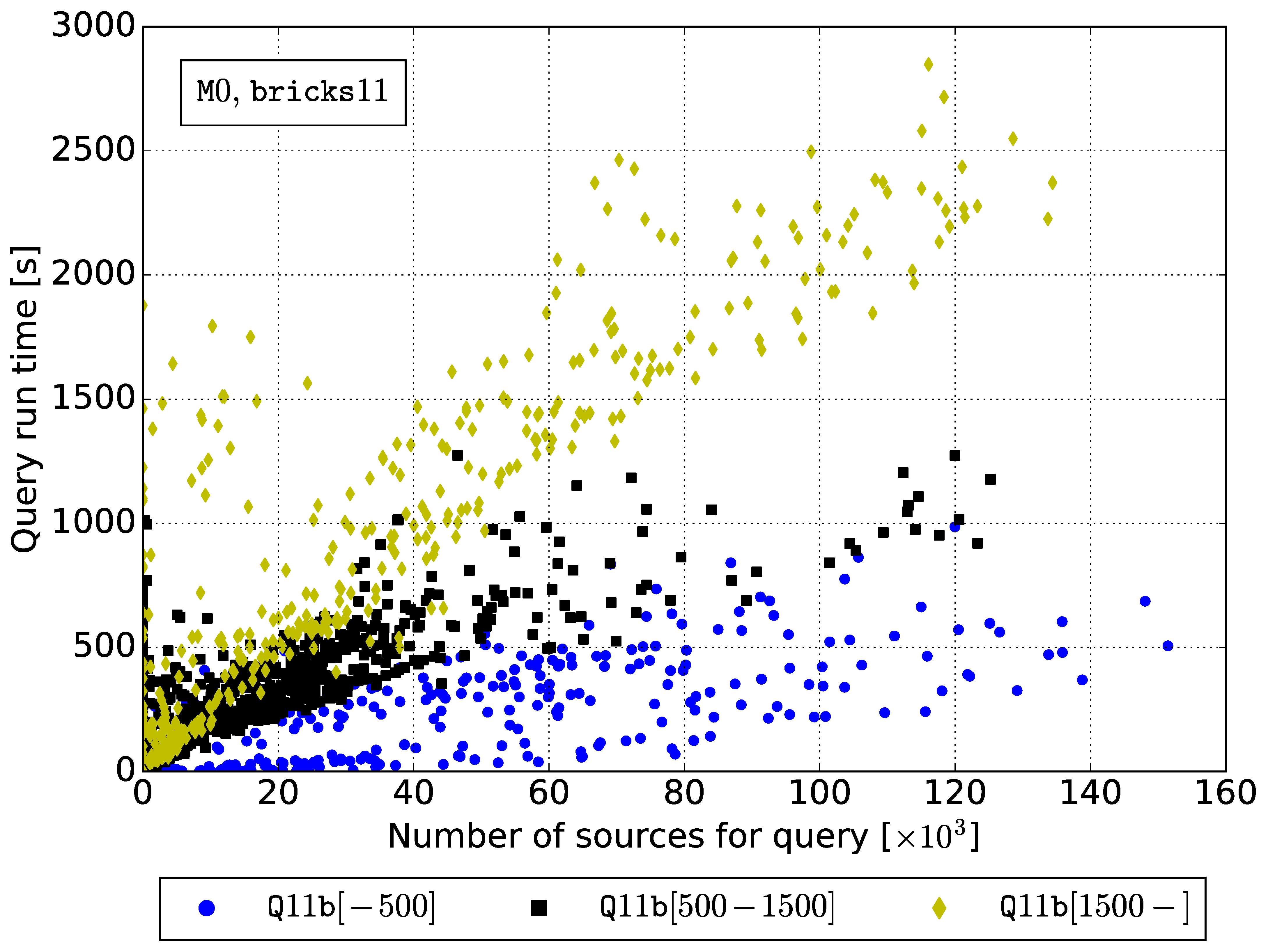}
\includegraphics[width=0.29\hsize,valign=t]{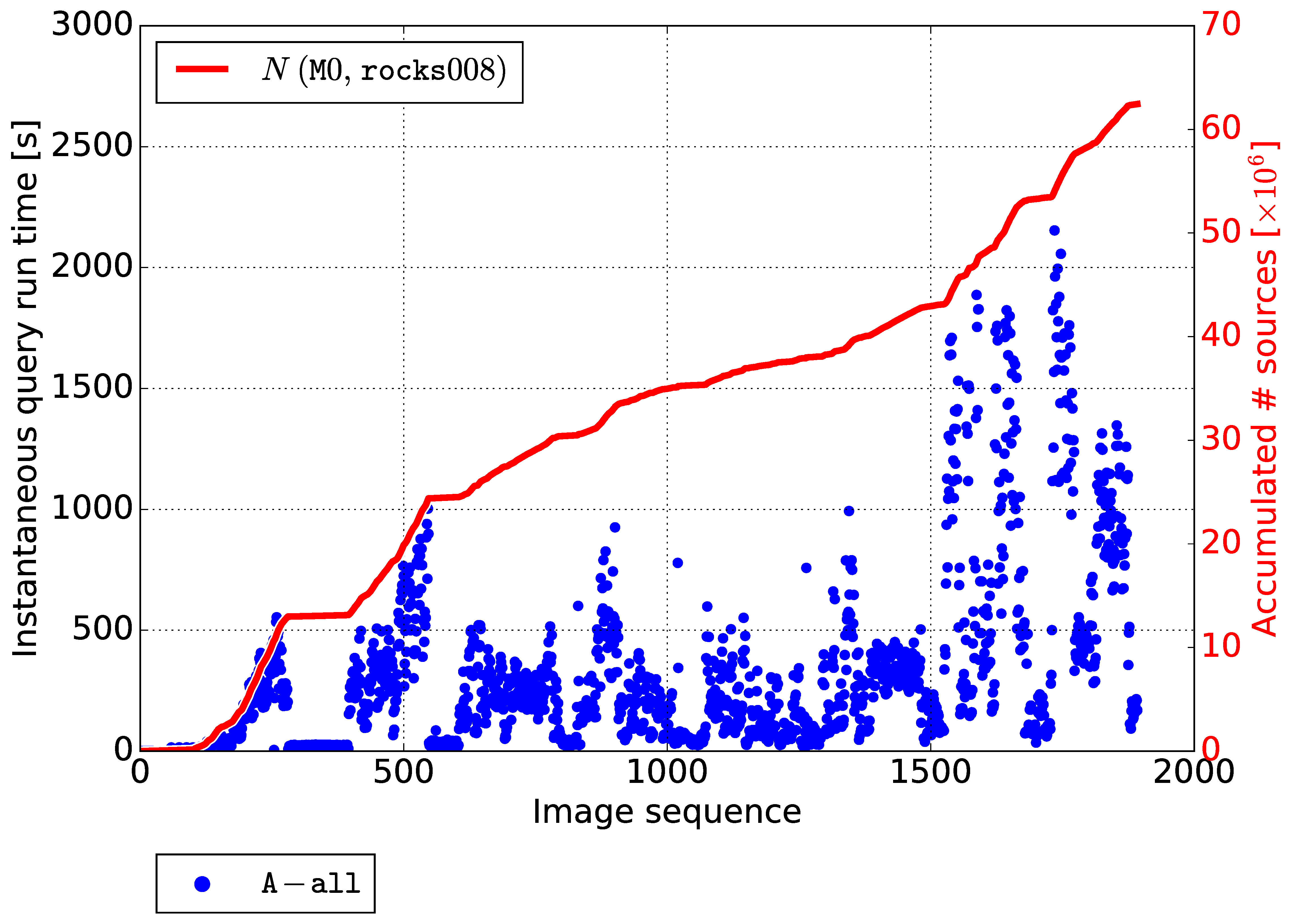}%
\includegraphics[width=0.29\hsize,valign=t]{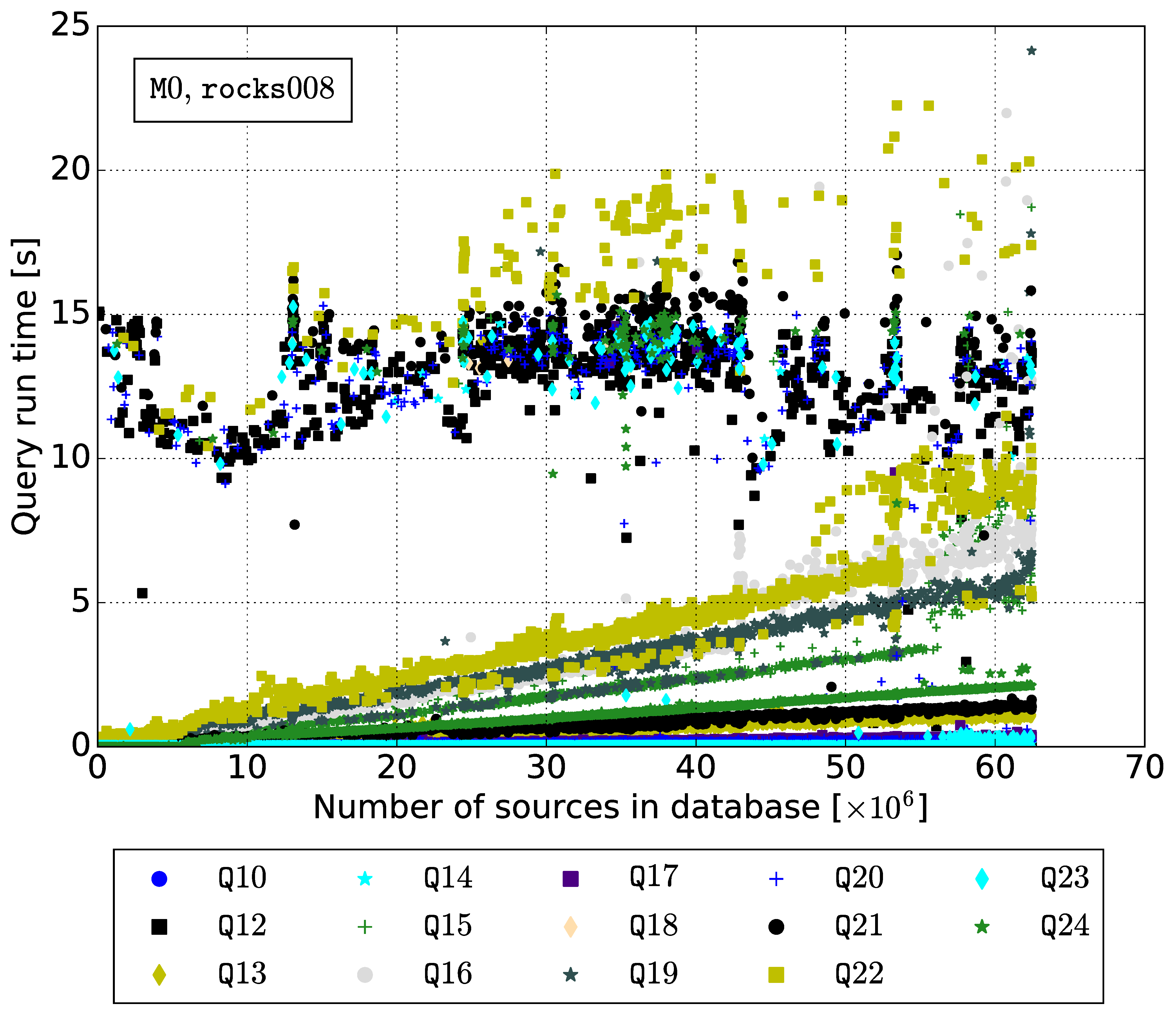}%
\includegraphics[width=0.29\hsize,valign=t]{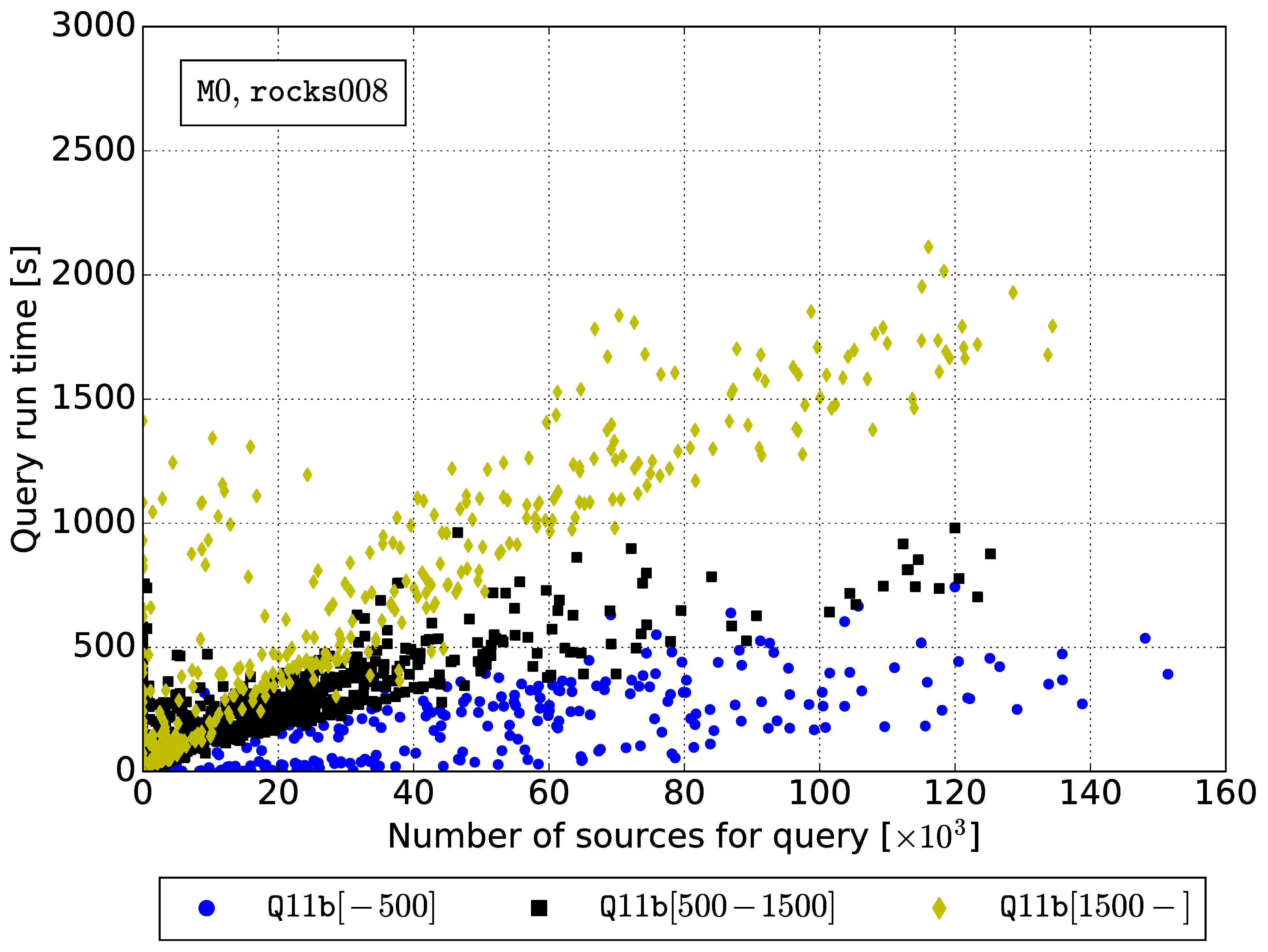}
\caption{
Query run times for cross-match module \texttt{M0}.
Per row the graphs show the run times for the different types of node
on which the pipelines ran; from top to bottom these are the 
\texttt{diamonds}, \texttt{stones}, \texttt{bricks} and \texttt{rocks}
nodes.
(See Table~\ref{tab:SciLconfig} for the node specifications.)
The left-column graphs show the summed run times of all queries
(left vertical axis) of module \texttt{M0} with respect to 
the image sequence number (horizontal axis).
The thick (red) line shows the actual accumulated 
number of sources stored in the database 
(right vertical axis) at the moment of \texttt{FITS} file processing.
The graphs in the middle column show the individual query run times 
according to the database size, 
which actually is the number of sources stored in the database
at the moment of query execution.
All but the cross-matching queries are shown.
The right-column graphs show the run times of the cross-matching query 
with respect to the number of rows in the query result set,
which approximates the number of entries in the source list.
The run times are divided into three parts,
where the image sequences run from start to 500, 
500 to 1500, and from 1500 up to the end.
}
\label{fig:a_M0}
\end{figure*}

The baseline cross-match module 
consists of fifteen queries in total of which 
the cross-matching query is labeled \texttt{Q11b}
(see \ref{app:q0}).
The graphs in Fig.~\ref{fig:a_M0} show the performance of
the cross-matching module as a whole (left column), 
the \emph{other} queries (middle column)
and the individual cross-matching query \texttt{Q11b} (right column),
while the rows specify the types of node
on which the pipeline ran.

The left-column graphs show the run times summed over
all fifteen queries versus the image sequence number.
During a pipeline run the total run time to process
a single \texttt{FITS} image fluctuates heavily and 
depends strongly on the growth rate of the database size.
Larger data input sizes will slow down the pipeline run.

From the right-column graphs it can be seen that the cross-matching query run times
are of the same order as the above mentioned summed query run times,
meaning that the pipeline run time
is determined by \texttt{Q11b}.
We have divided the cross-matching query run times in these graphs into 
three consecutive parts according to their image sequence number:
the first 500 images, the middle part of images 500--1500 and 
the remaining images starting from sequence number 1500.
This points out that cross matching is not only a function of number of query rows, 
but also of database size, 
since for the same number of rows
the query run times slow down as 
the database fills up with sources.
Searching for counterpart candidates takes longer, since larger tables have to be scanned.

Comparisons of the run times of the cross-matching query to the other queries,
shown in the graphs in the middle column of Fig.~\ref{fig:a_M0},
reveal a difference of more than two orders of magnitude, 
making the cross-matching query the most dominant one in module \texttt{M0}.
The middle graphs show that most queries evolve linearly in time 
as the database size grows, 
while only a few run in constant time independent of the database size.

It all indicates that 
the cross-matching query and thus the baseline module scales with database size
in $\mathcal{O}(NM)$ time, where $N$ is the number of sources
in the database and $M$ the number of sources in the resulting candidate list,
which is approximately equal to the number of entries 
in the source list originating
from the \texttt{FITS} file.

The pipeline performance on the \texttt{diamonds} and
\texttt{bricks} nodes are similar,
because for the former the RAM size is large enough
and for the latter RAM size is sufficient
in combination with the low latency storage access 
of the SSDs.
On the \texttt{stones} and \texttt{bricks} nodes
some of the \emph{other} queries are memory bound.
The fast CPU of the \texttt{rocks} node does not compensate 
its small RAM size,
which effect is more prominent 
when the database size is larger.
In these cases the operating system starts memory swapping data
from RAM to disk,
a process that impacts the pipeline performance negatively.
The larger storage access latency for HDDs as compared to SSDs
makes the pipeline runs slower and less smooth on a \texttt{stones} node
than on a \texttt{bricks} node, despite its faster CPU.

\begin{figure*}[!p]
\centering
\includegraphics[width=0.29\hsize,valign=t]{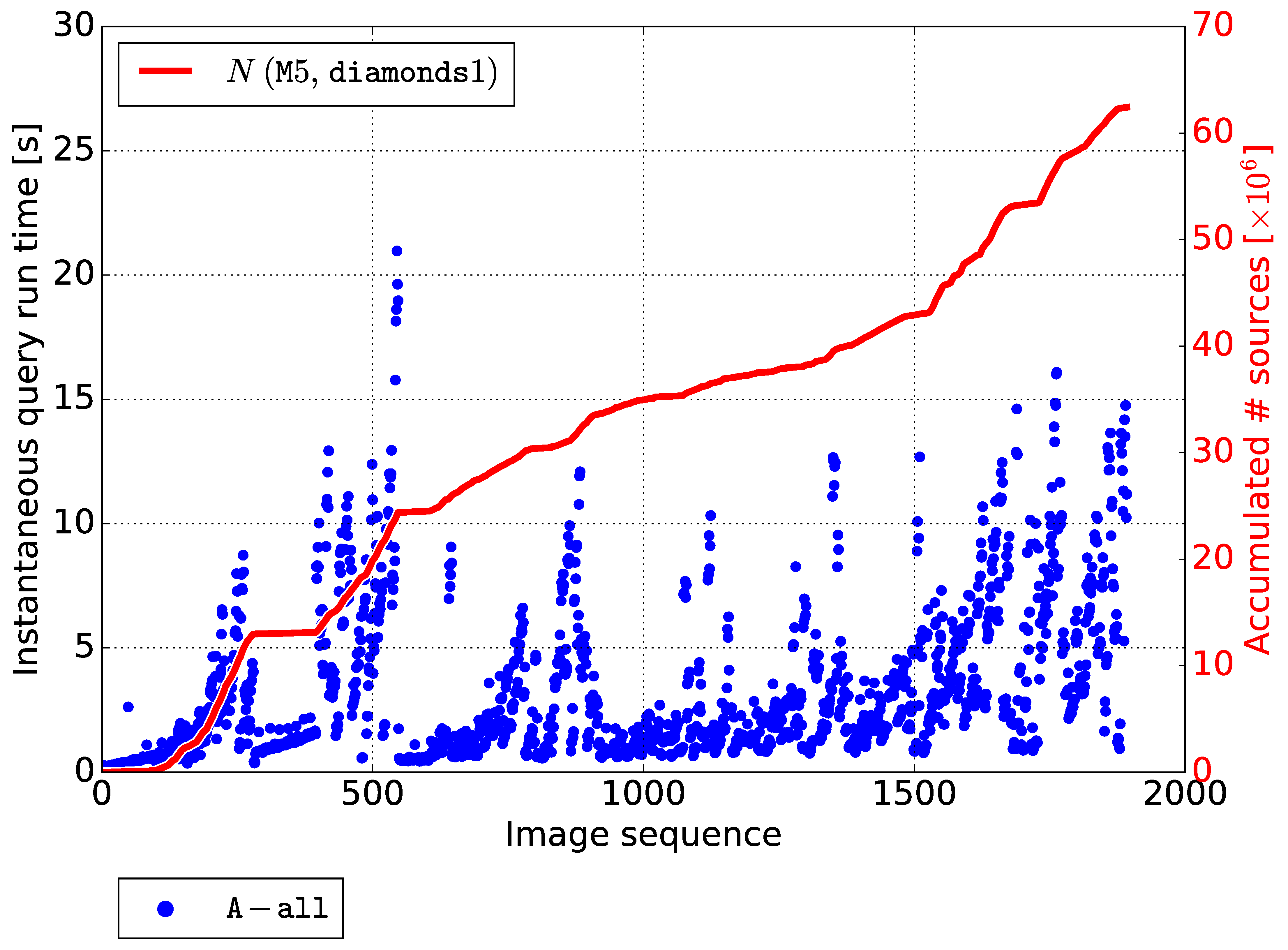}%
\includegraphics[width=0.29\hsize,valign=t]{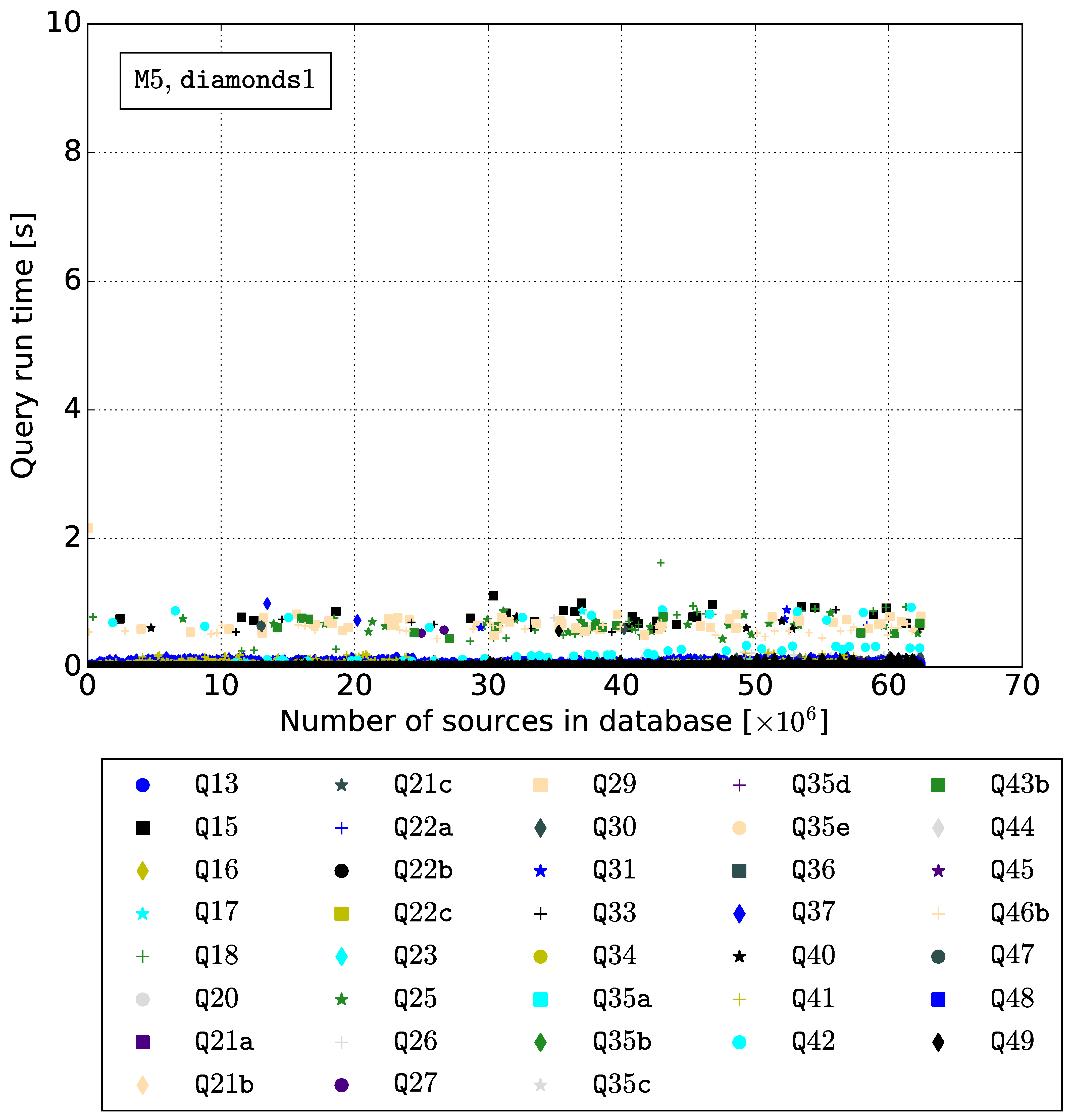}%
\includegraphics[width=0.29\hsize,valign=t]{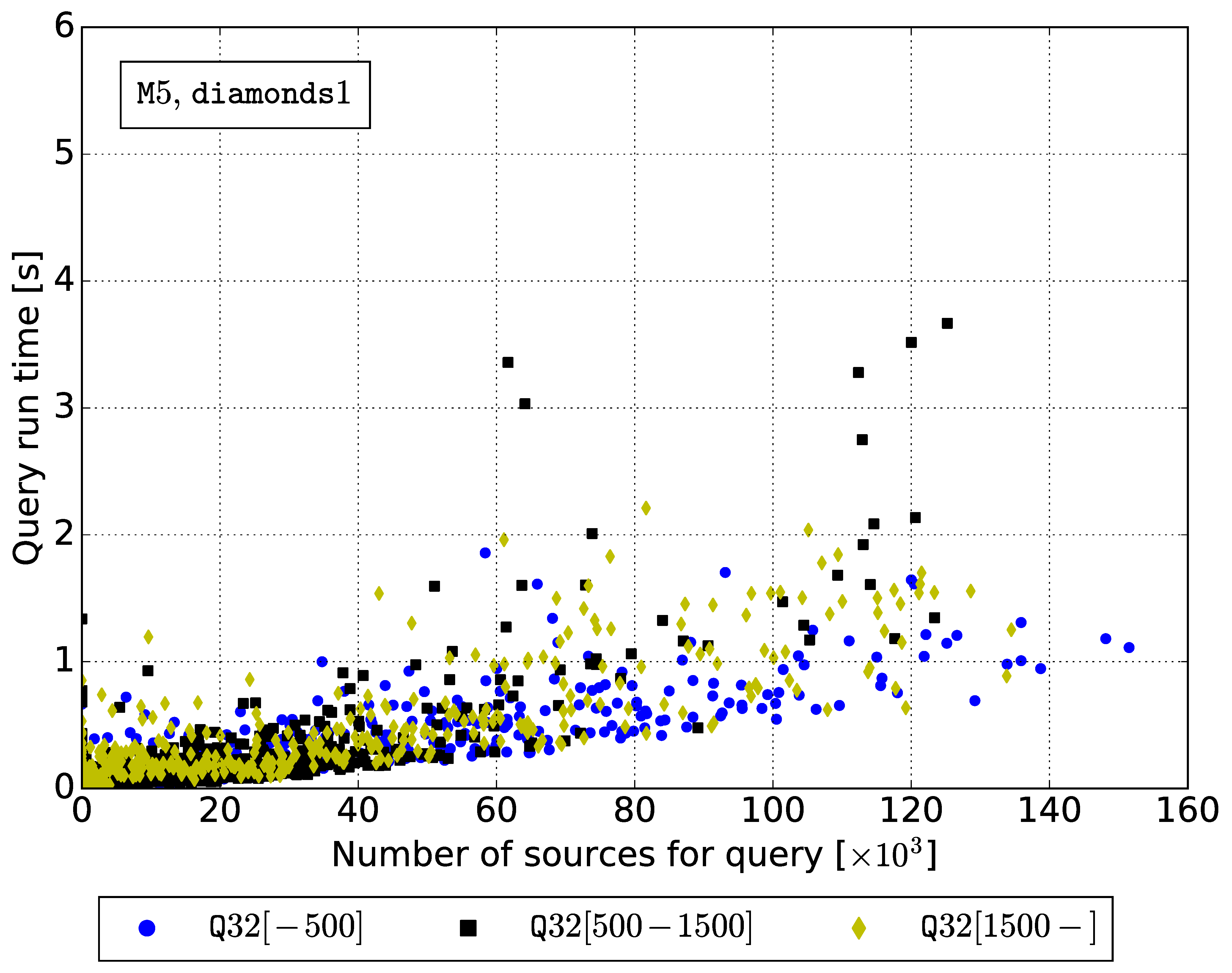}
\includegraphics[width=0.29\hsize,valign=t]{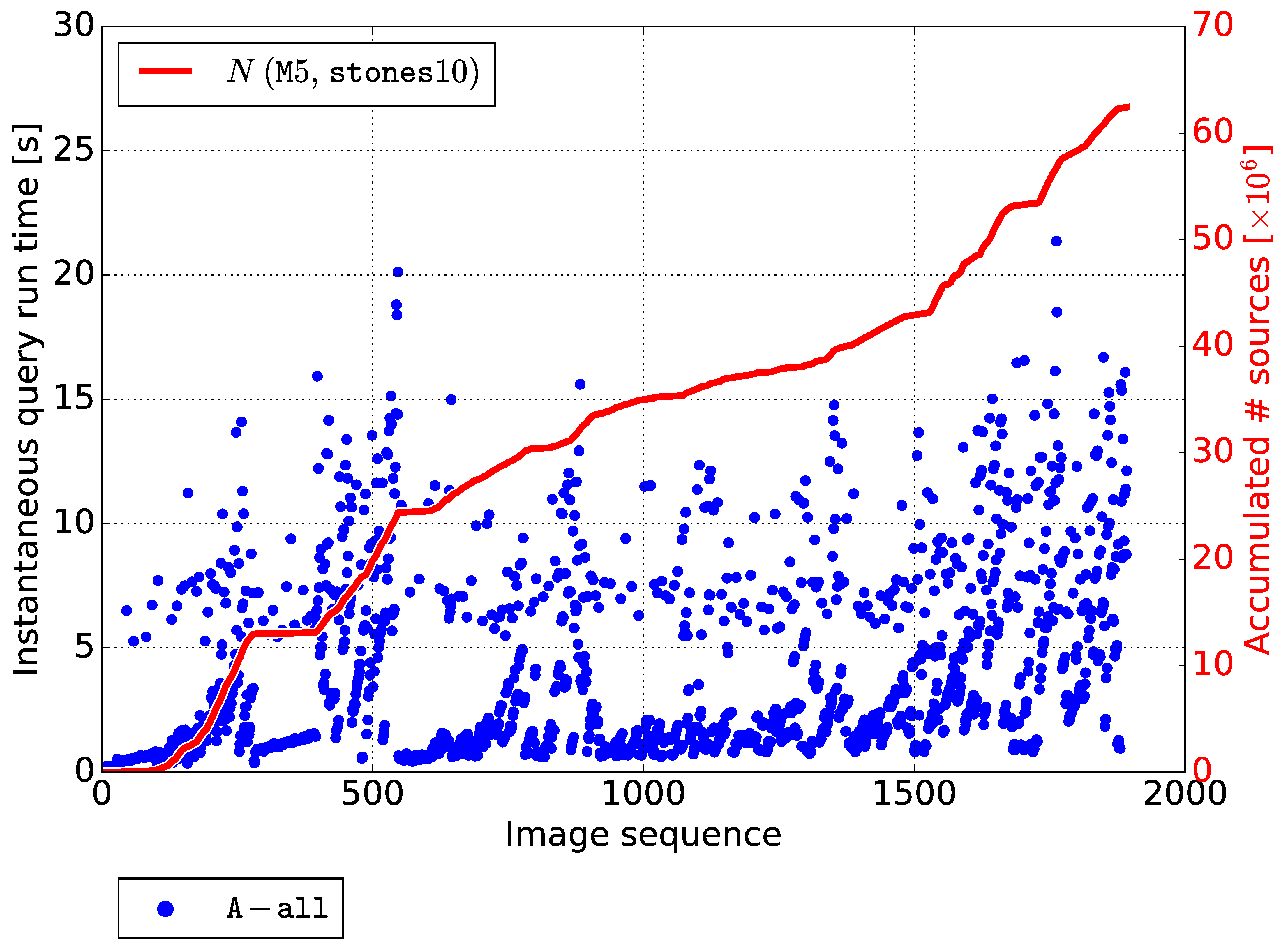}%
\includegraphics[width=0.29\hsize,valign=t]{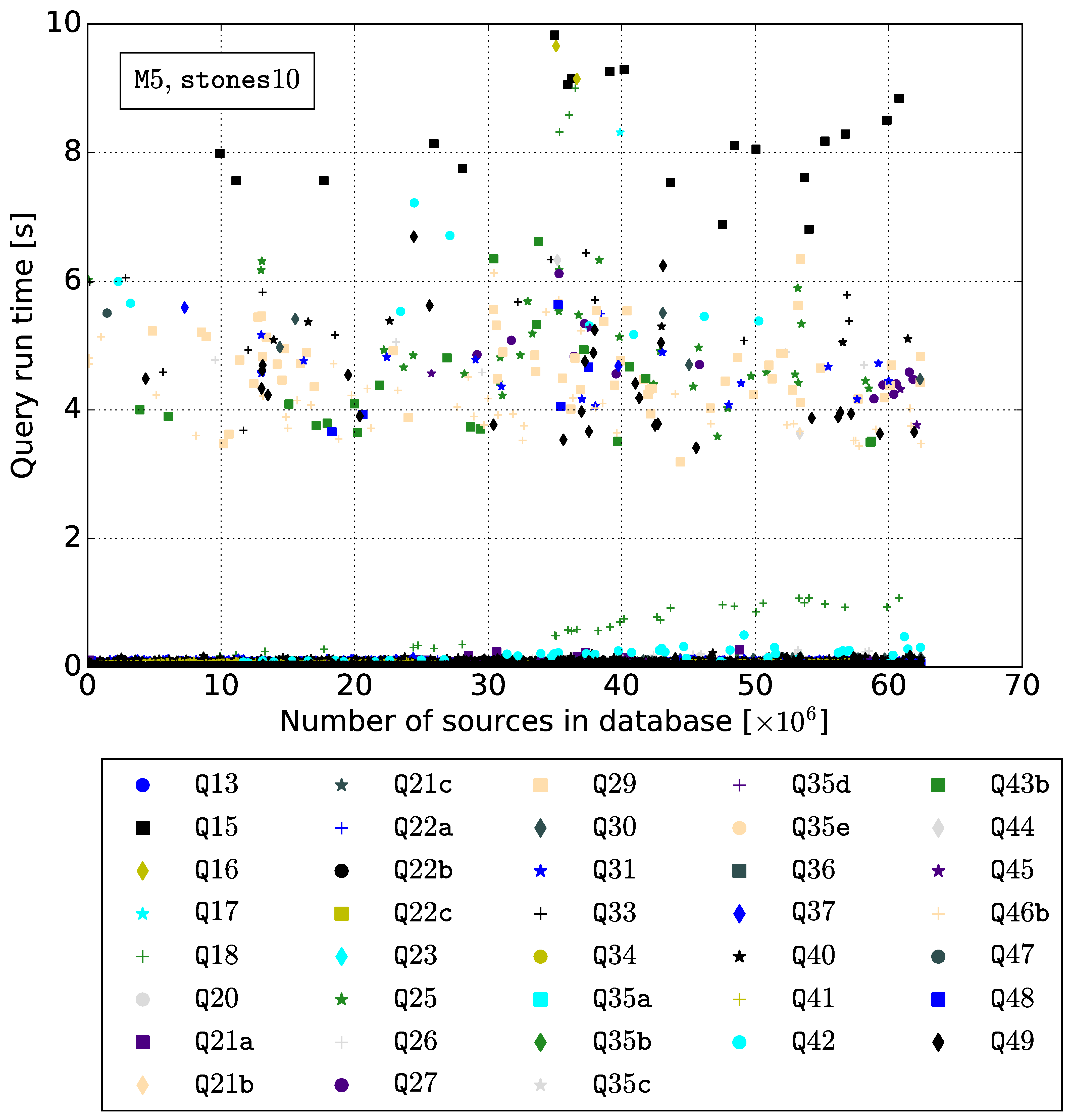}%
\includegraphics[width=0.29\hsize,valign=t]{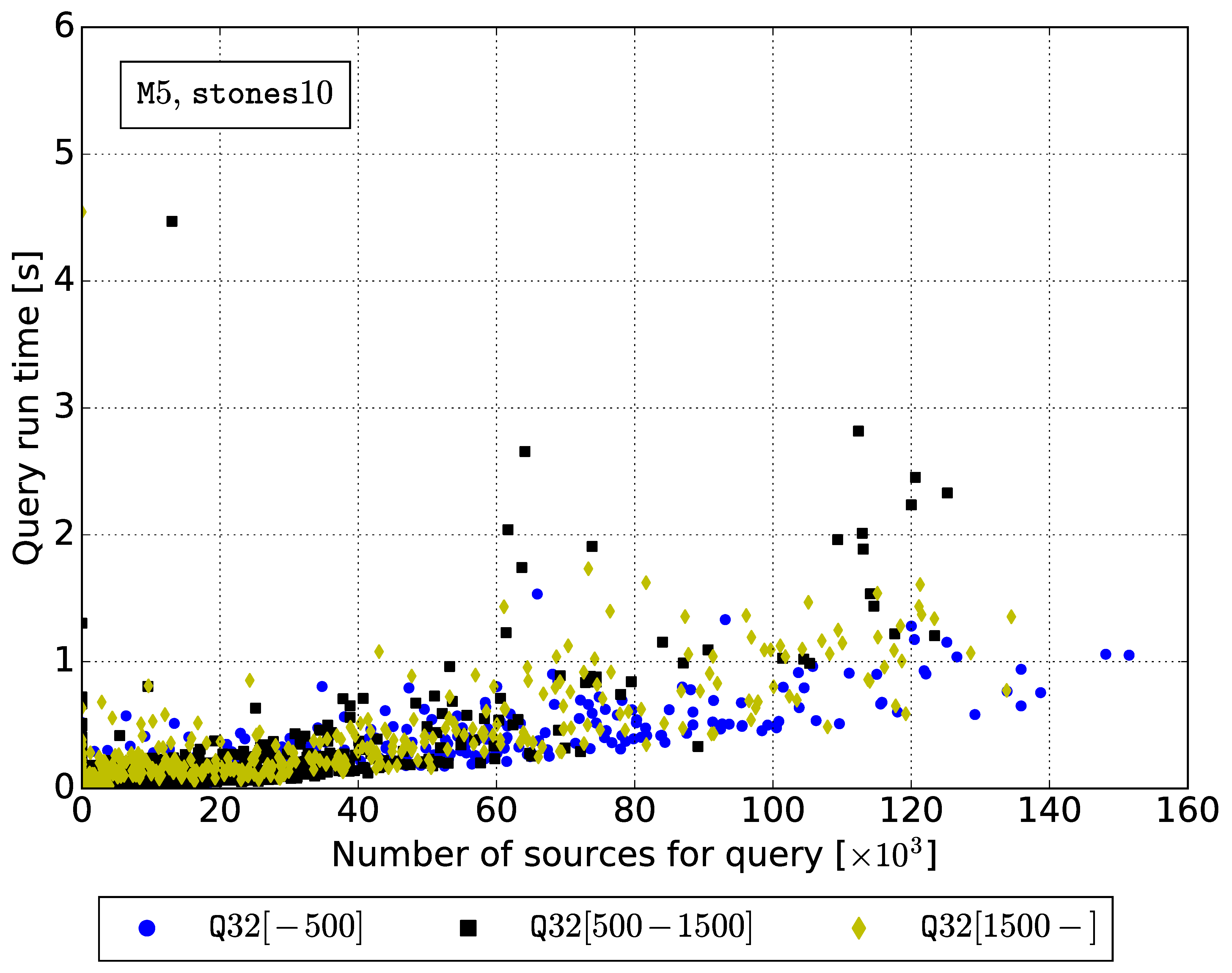}
\includegraphics[width=0.29\hsize,valign=t]{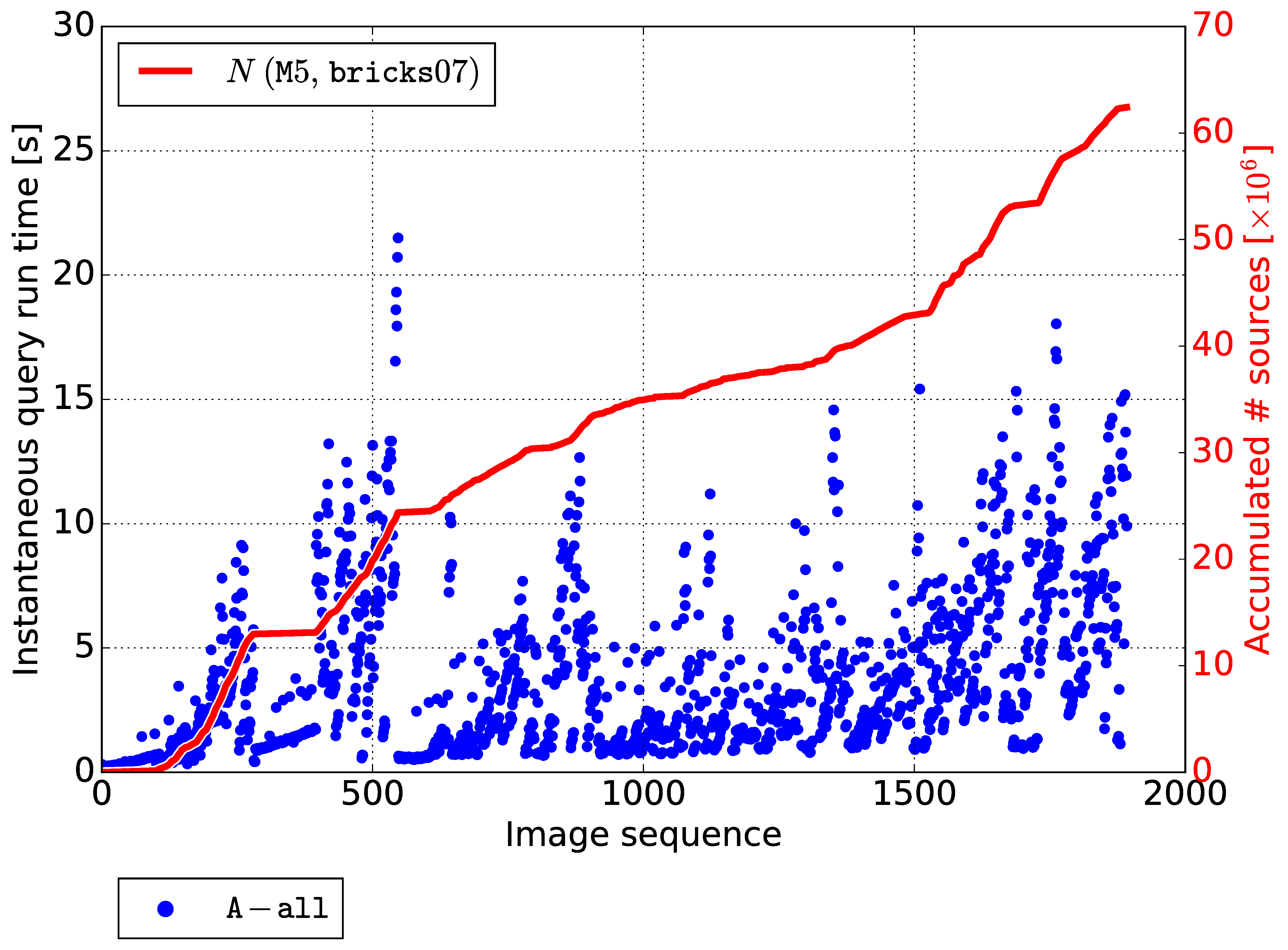}%
\includegraphics[width=0.29\hsize,valign=t]{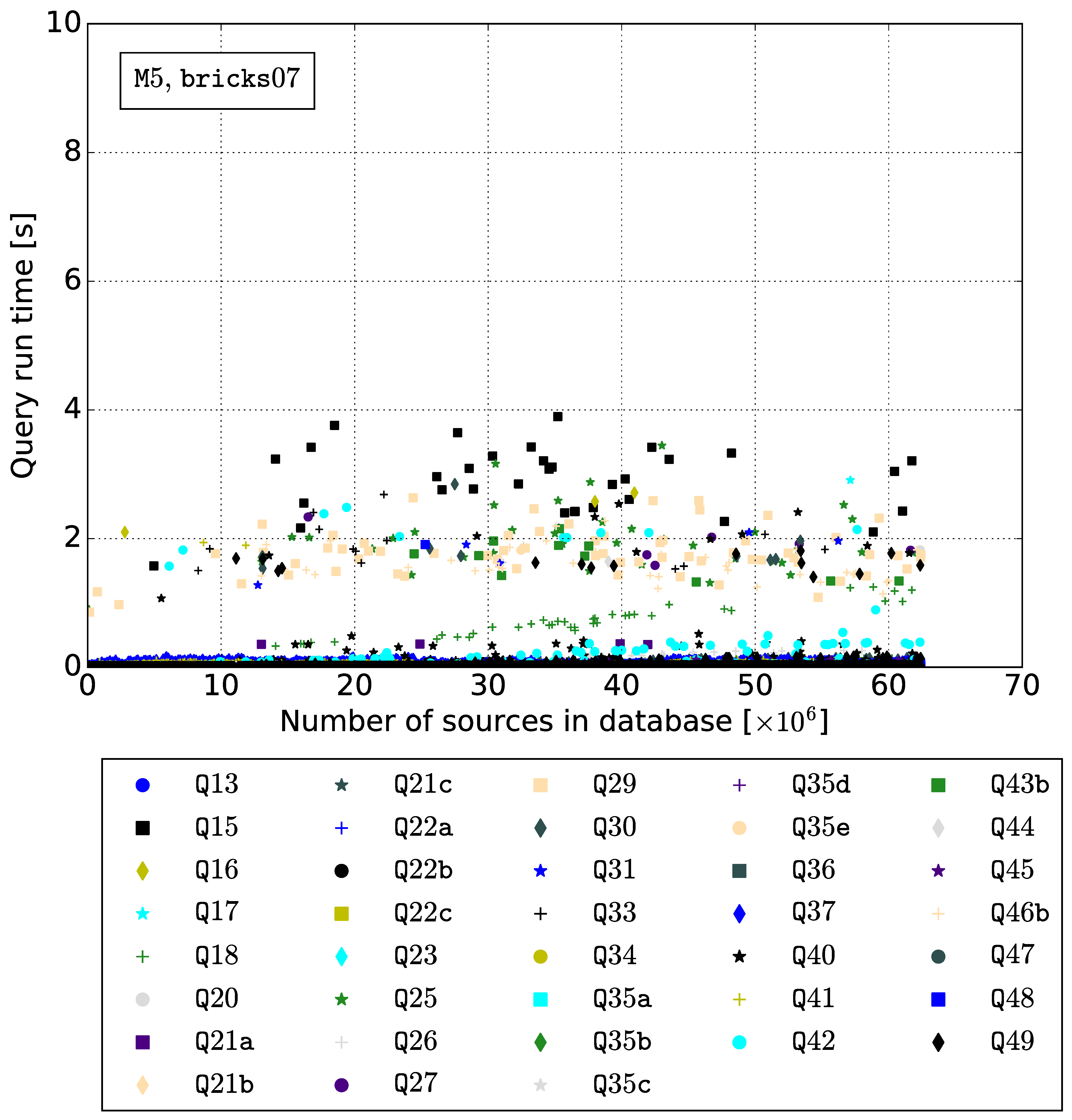}%
\includegraphics[width=0.29\hsize,valign=t]{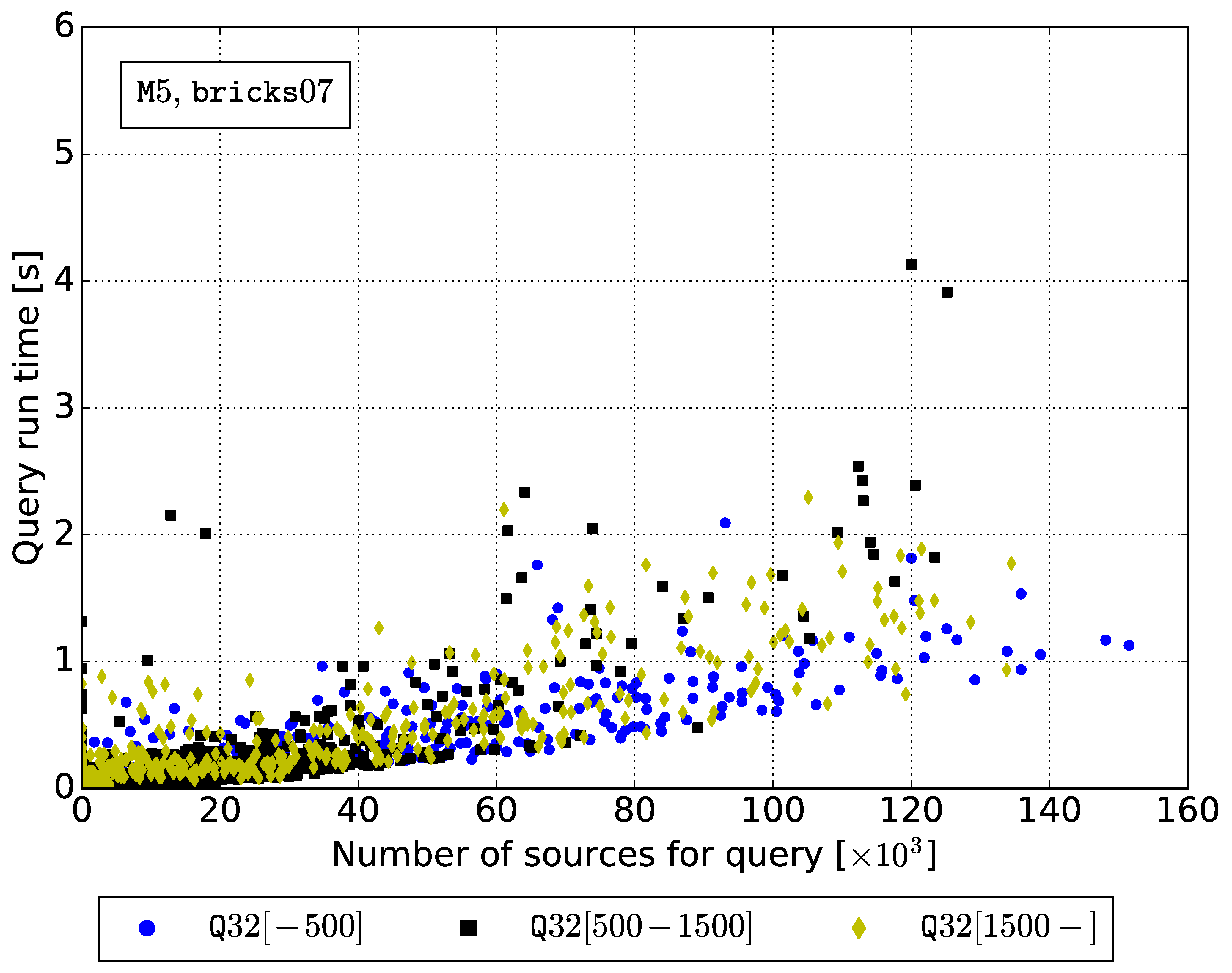}
\includegraphics[width=0.29\hsize,valign=t]{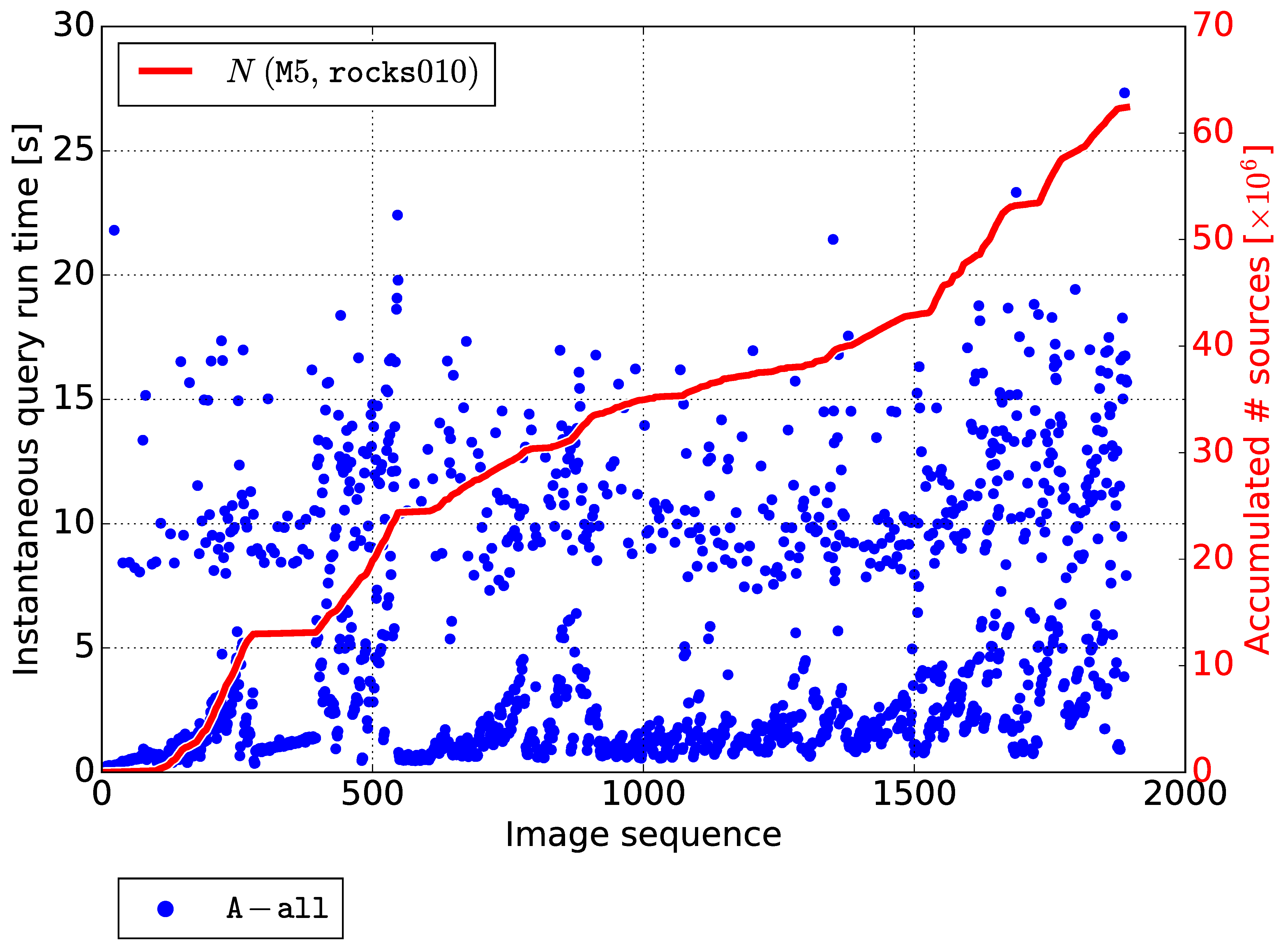}%
\includegraphics[width=0.29\hsize,valign=t]{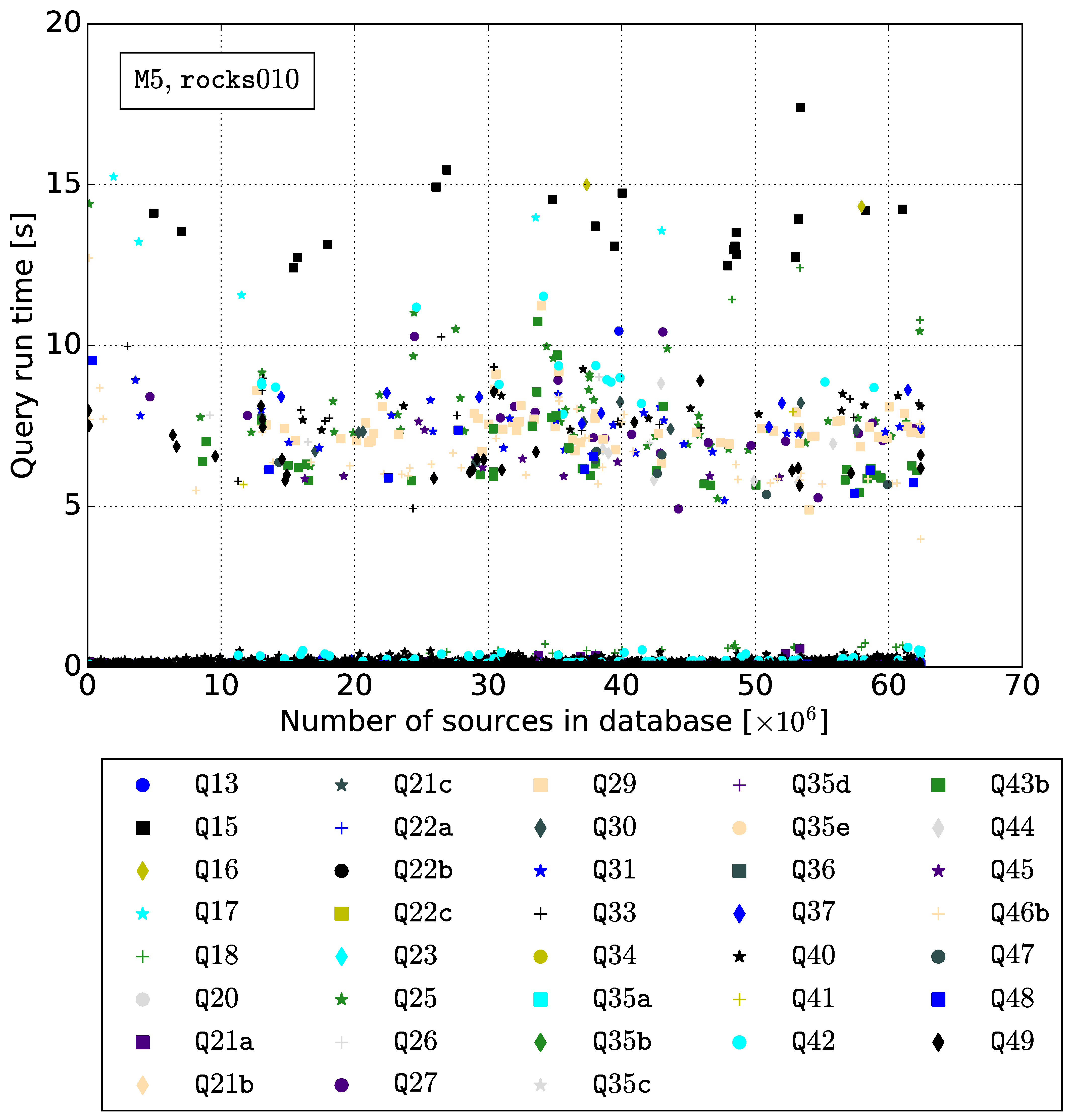}%
\includegraphics[width=0.29\hsize,valign=t]{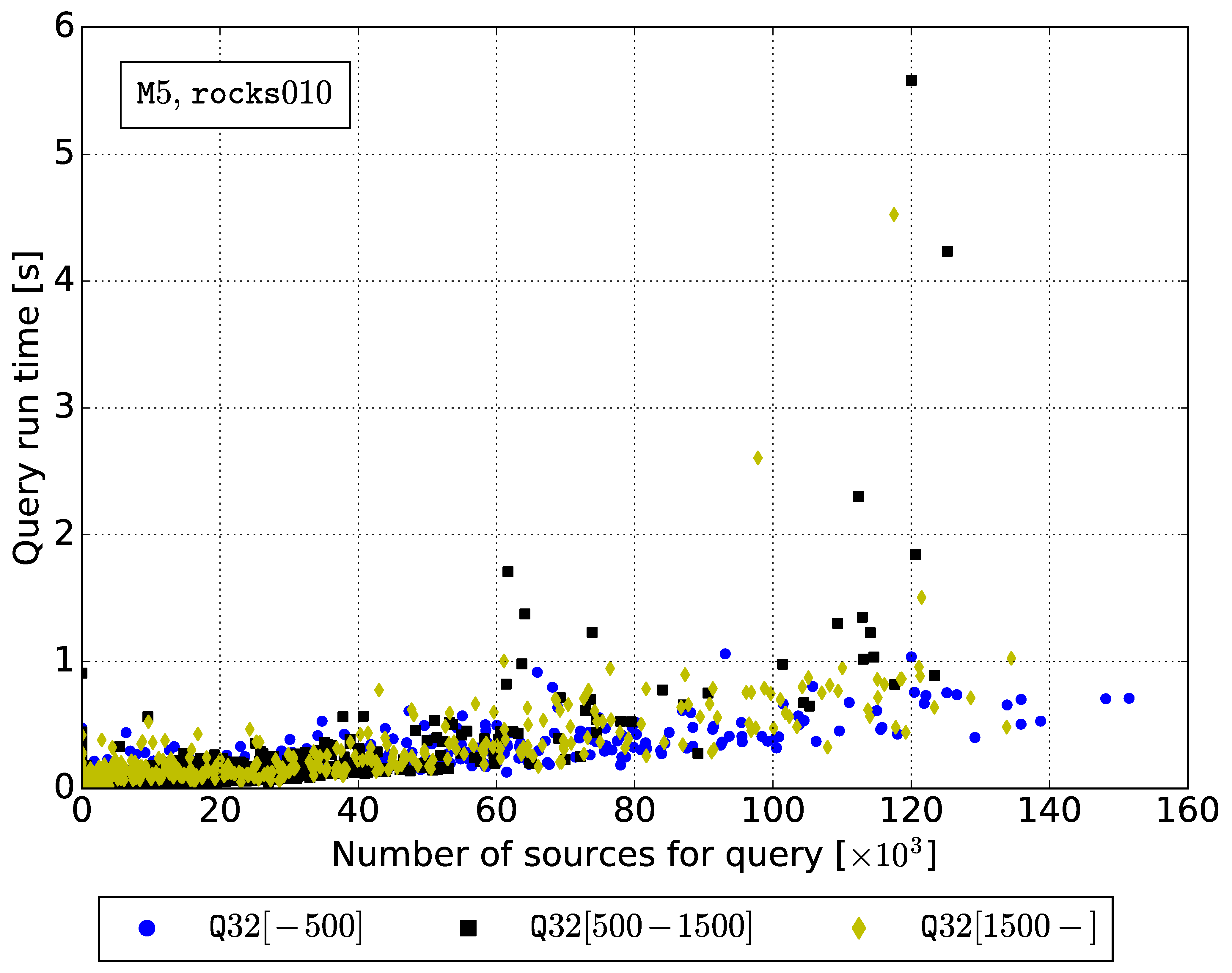}
\caption{%
Query run times for cross-match module \texttt{M5}.
The left-column graphs show the summed run times of all queries 
vs.\ the image sequence number. 
The graphs in the right column show the cross-matching query
run times with respect to the number of sources in the database
($N$)
and the remaining queries that run in constant time 
with respect to $N$ 
are shown in the middle column.
(See caption of Fig.~\ref{fig:a_M0} for further 
descriptions of the graphs.)
The performance of the remaining queries of the module are depicted
in Figs.~\ref{fig:a_M5_max} and \ref{fig:a_M5_lin}.
}
\label{fig:a_M5_const}
\end{figure*}

\subsubsection{The partitioned cross-matching module \texttt{M5}}
\label{sec:res_M5}

\begin{figure*}
\centering
\includegraphics[width=0.3\hsize,valign=t]{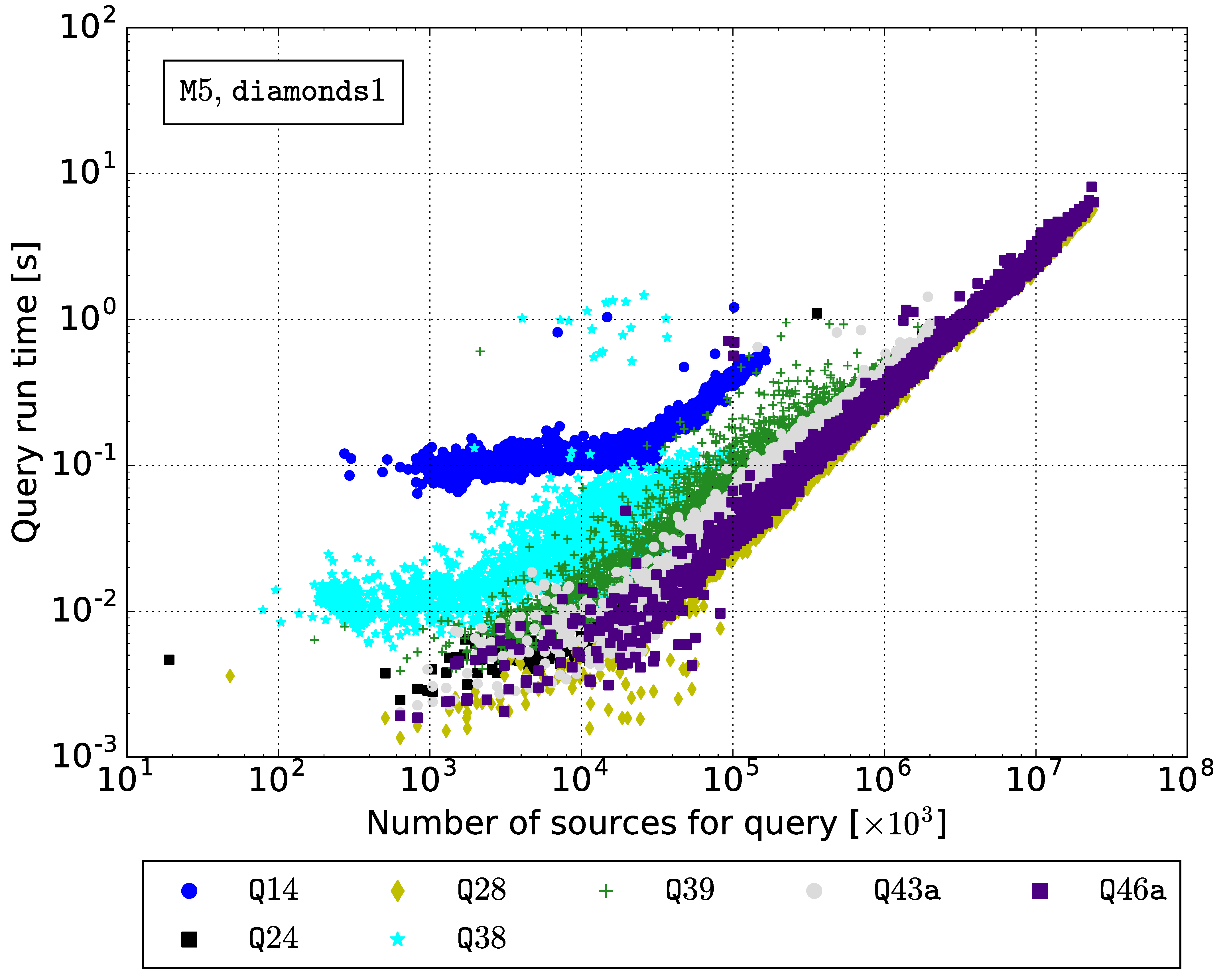}%
\includegraphics[width=0.3\hsize,valign=t]{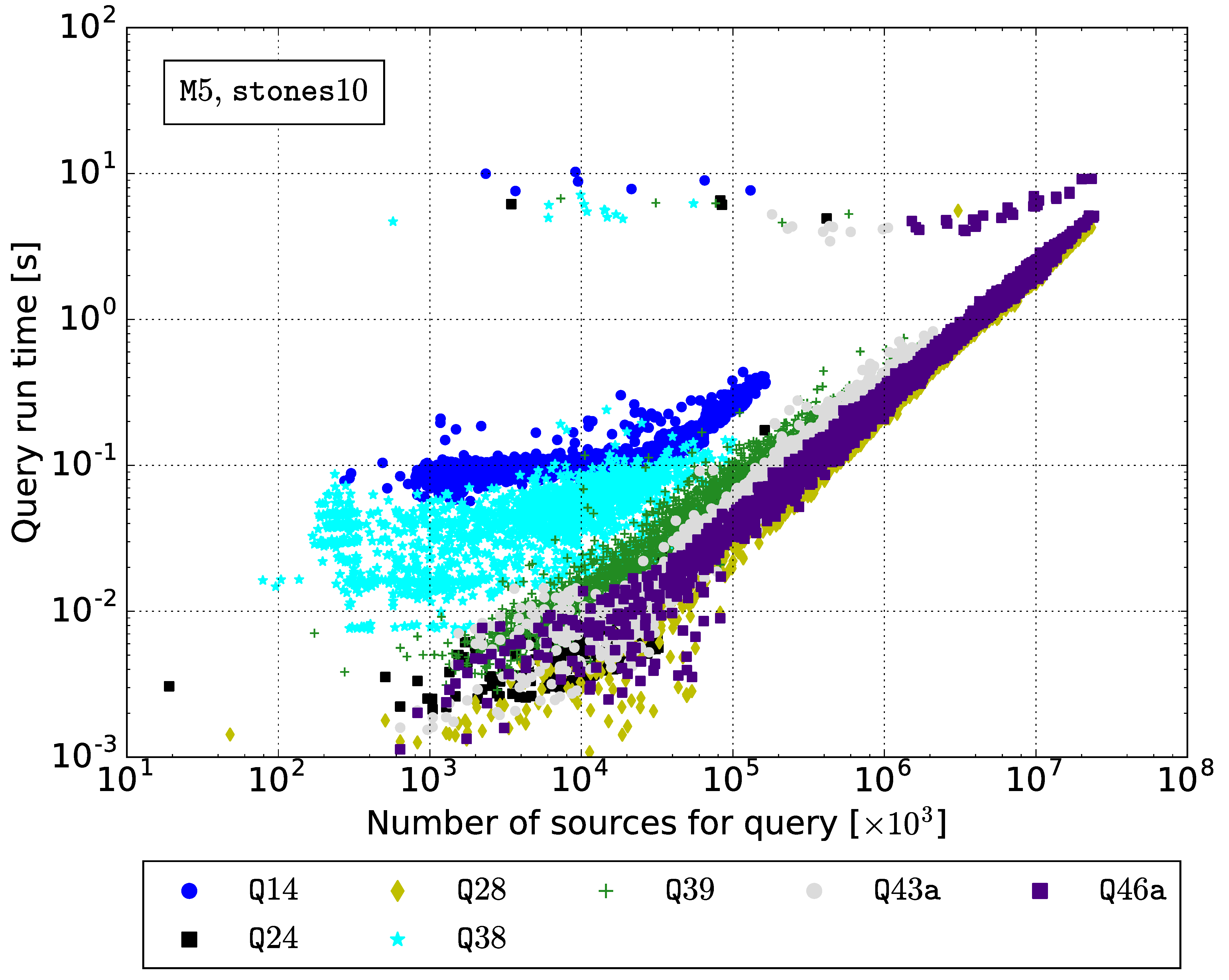}\\
\includegraphics[width=0.3\hsize,valign=t]{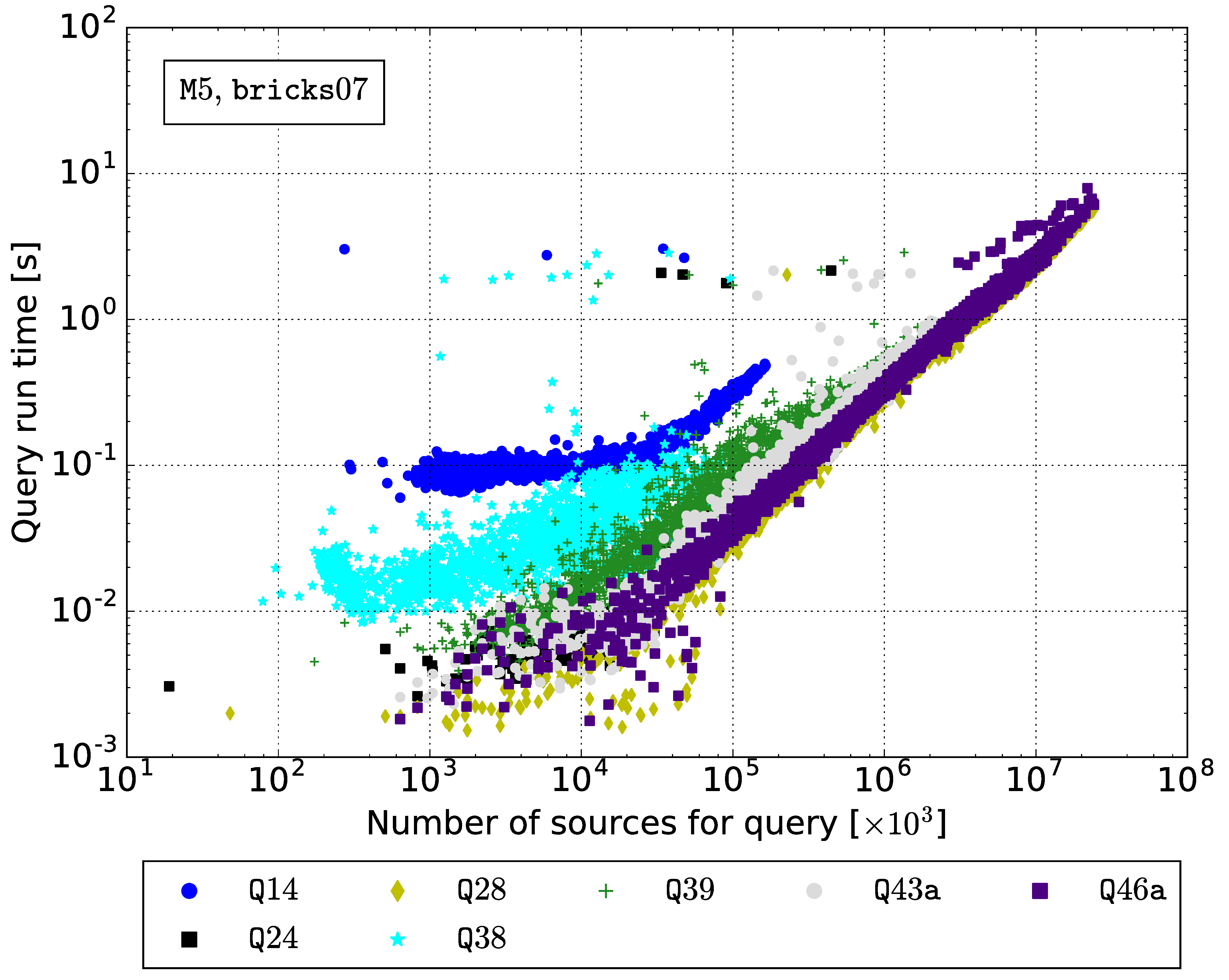}%
\includegraphics[width=0.3\hsize,valign=t]{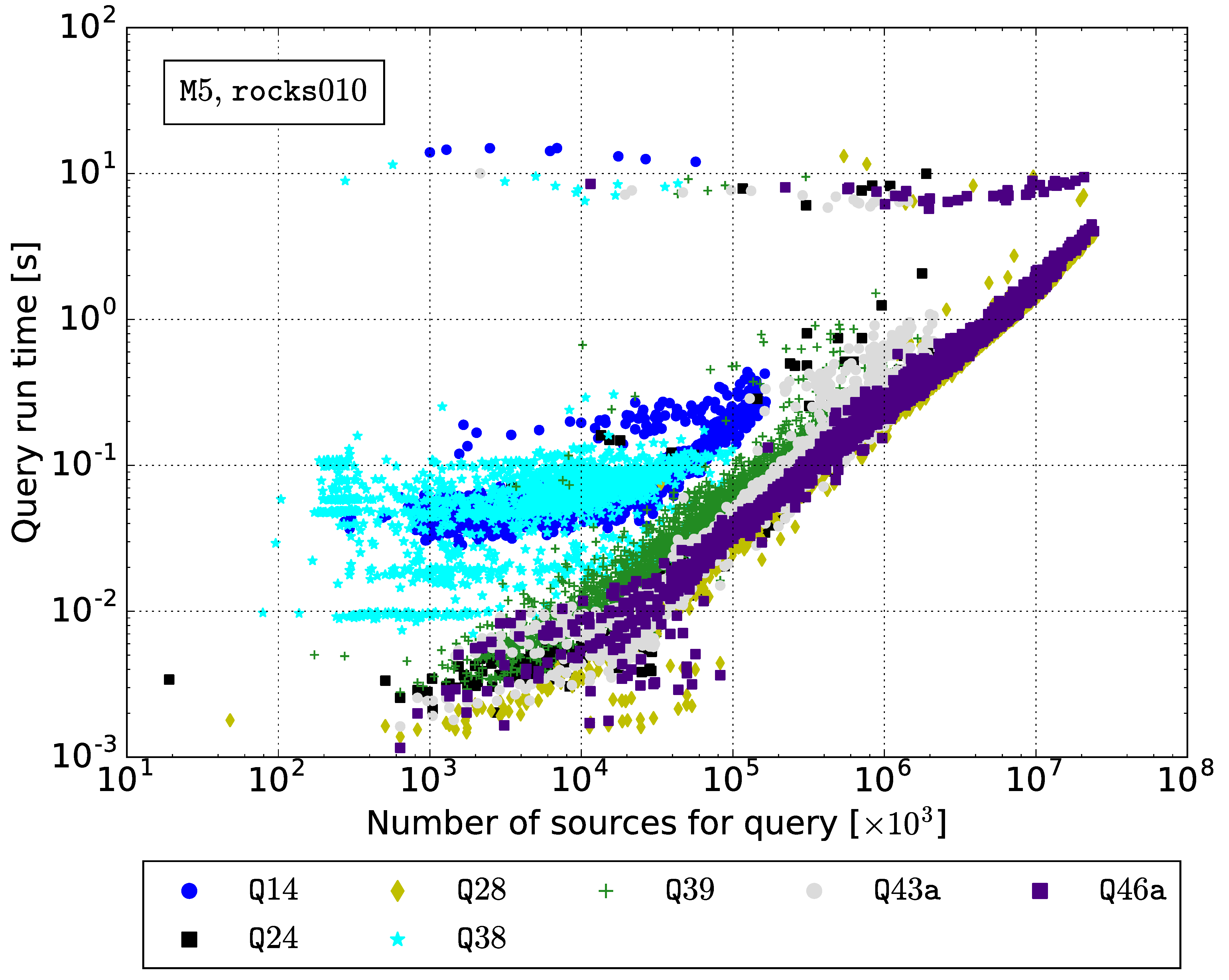}
\caption{%
Module \texttt{M5} queries that run in linear time.
The plots show the remaining queries that
depend on the number of rows,
which approximates the source list size.
Pipeline runs were on the different cluster nodes.
}
\label{fig:a_M5_lin}
\end{figure*}

In module \texttt{M5} the cataloged sources are distributed 
over multiple tables according to their declination zone
as described in \S~\ref{sec:descr_M5}.
The number of rows a query touches
is now limited by the size of the partitioned tables,
which is for zone widths of 1~degree 
two orders of magnitude smaller than the 
unpartitioned version.
Query run times now only depend either
on the number of rows of partitioned tables or on the size of the result sets
instead of the size of the entire database.
This is nicely demonstrated in 
Figs.~\ref{fig:a_M5_const} and \ref{fig:a_M5_lin},
where the run-time performance of all queries in
module \texttt{M5} are shown.

The cross-matching query, labeled \texttt{Q32}
in module \texttt{M5}, is shown in the right-column
graphs of Fig.~\ref{fig:a_M5_const}.
Because the partitioning restricts the number of sources,
the query run times do not depend on the database size anymore.
Therefore, in contrast with the 
baseline
module,
the cross-matching run times do not hinge on 
the image sequence number, 
but are determined by the 
number of entries in the result set, 
i.e.\ the number of counterpart candidates,
which is approximately equal to the number of entries
in the source list.
This makes the \texttt{M5} cross-matching query to run 
stable and faster over time than its counterpart in module 
\texttt{M0}.

The summed query run times fluctuate irregularly
over the course of images, 
as can be seen in the left column of Fig.~\ref{fig:a_M5_const}.
This originates from query contributions that depend 
mainly on the input source-list size,
which varies from image to image,
but the run times never exceed the cadence time.
The middle graphs in Fig.~\ref{fig:a_M5_const} show all queries
that run in constant time.
It can be seen that for the different nodes
the scatter of the run times 
increases with decreasing RAM size.
This is caused by queries that run in 
linear-time complexity and 
compete with all queries for the same memory resources.
Queries that depend on the number of rows need to 
allocate a relatively larger percentage of total memory
on the nodes with smaller RAM sizes,
leaving a smaller amount to the remaining queries
which in turn results in longer query run times.

\begin{figure}
\centering
\includegraphics[width=0.7\hsize]{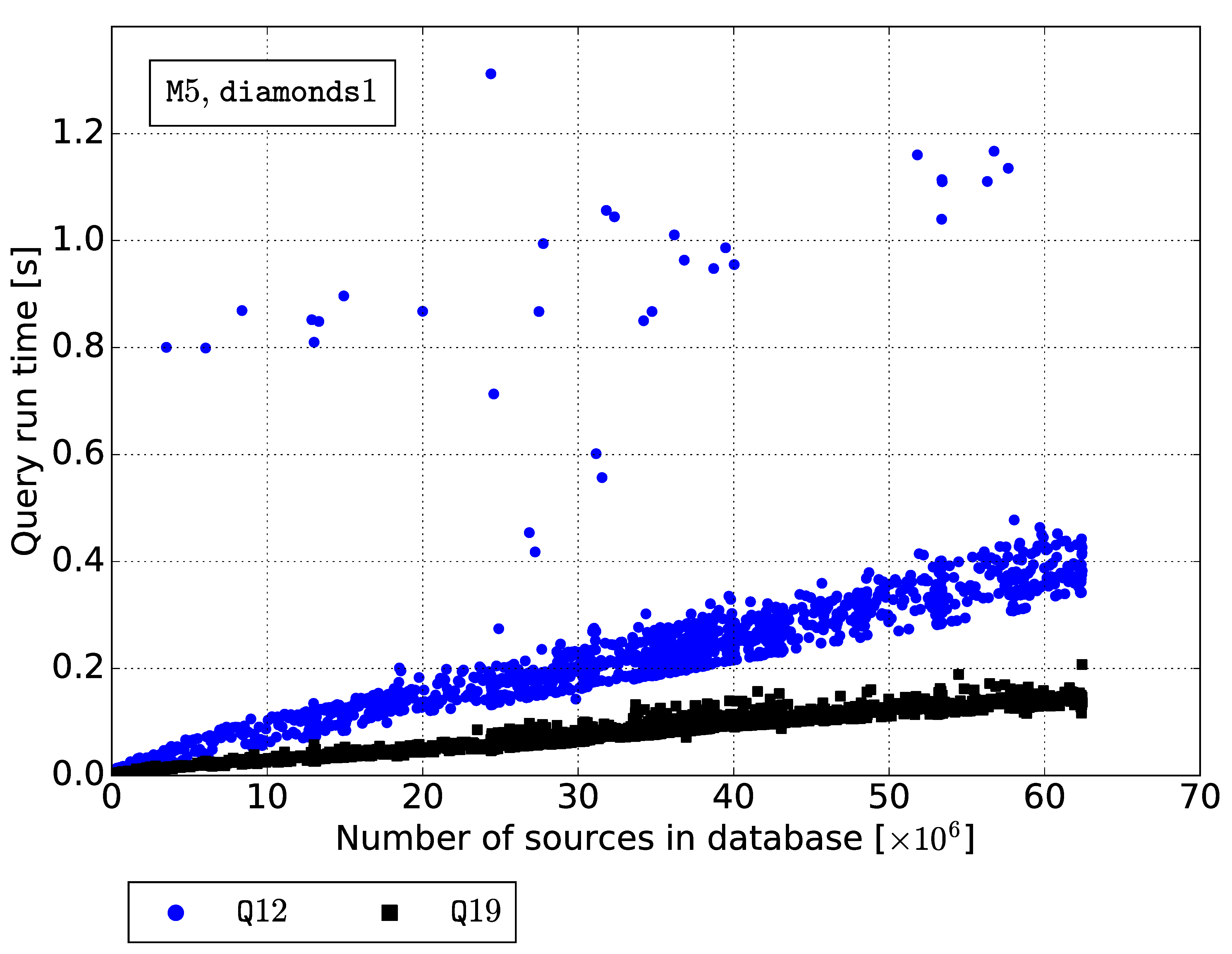}
\caption{%
Run time vs.\ number of sources in the database 
of the two queries that call the 
SQL \texttt{MAX} function in module \texttt{M5}
on the \texttt{diamonds} node.
}
\label{fig:a_M5_max}
\end{figure}

The small group of queries that scales linearly with the number of rows
are displayed in Figs.~\ref{fig:a_M5_max} and \ref{fig:a_M5_lin}.
Fig.~\ref{fig:a_M5_max} shows two queries of which the 
run time increases linearly with respect to the database size.
The respective queries select the maximum \texttt{ID} value of the 
growing tables of extracted (\texttt{Q12})
and cataloged sources (\texttt{Q19}).
Note that the latter is a \texttt{MERGE} table
of many partitions.
Since the number of unique sources is smaller than
the total number of extracted sources 
the slope of query \texttt{Q19} is less steep.
Although it is not noticeable in these experiments, 
but if we extrapolate the database size to larger numbers,
the run times will hit the cadence time at some point.
A fairly easy way to correct this part to run in constant time 
would be to maintain the values at the application level.
However, a solution where the database itself keeps track of 
statistical parameters, e.g., average, minimum, maximum,
standard deviation, is more elegant and is considered in future releases.

The graphs of Fig.~\ref{fig:a_M5_lin} show the 
\texttt{M5} query run times that 
do not run in constant time.
The areas of the query run times are confined to 
certain regions. 
Queries \texttt{Q46a} and \texttt{Q28} are working on the light-curve table
and take into account an order of magnitude more entries than
the queries that run over the catalog table with unique sources (\texttt{Q43a} and \texttt{Q24},
where the latter is hidden behind the markers of other query run times).
All four query run times depend on the number of sources
in the partitions, i.e., the source density.
Numbers of new datapoints appended to the light-curve table (\texttt{Q39})
do not exceed $10^6$, 
whereas new entries for the extracted source table (\texttt{Q14}) 
and catalog table (\texttt{Q38})
both do not exceed the number of source list entries.
The graphs show that the performance is controlled and that the queries run at
acceptable speeds.

\subsection{Overall pipeline runtime}
\label{sec:results_overall}

\begin{table*}[!t]
\centering
\begin{tabular}{cl||S[table-format=3.1]|S[table-format=3.1]|S[table-format=3.1]|S[table-format=3.1]}
\hline
module &  & \textbf{diamonds} & \textbf{stones} & \textbf{bricks} & \textbf{rocks} \\
\hline
\hline
\texttt{M0} & pipeline & 659.0 & 575.1   & 750.6 &  589.4 \\
            & query    & 658.4 & 572.7   & 749.8 &  585.6 \\
\hline
\texttt{M5} & pipeline &   9.6 &  14.8 &  12.0 &   23.2 \\
            & query    &   6.5 &   7.8 &   7.0 &   10.0 \\
\hline
\end{tabular}
\caption[]{Accumulated processing times (in ks) of
the modules for a complete pipeline run,
specified per type of node.
Node rows give the specific node on which the pipeline ran,
pipeline rows specify the total pipeline run time and 
query rows report the total run times of all queries summed.
}
\label{tab:overall_ortho}
\end{table*}

Table~\ref{tab:overall_ortho} reports the total time of the pipeline runs and 
the summed execution time of all queries, i.e.\ loading and cross-matching, for the 
baseline and the alternative module,
specified per cluster node.
It shows that absolute increment of the pipeline overhead is
more prominent in the alternative module \texttt{M5}
than in the baseline module \texttt{M0}.
This is caused by fact that we do not include the query commit times in the
query runtime results.
The larger number of queries and thus commits in the alternative module contribute more
to the total pipeline runtimes.

The scalability of the baseline module \texttt{M0}, 
which extrapolates LOFAR's TraP pipeline from the radio to the optical domain, 
is limited.
Most queries write data, 
i.e.\ append or update,
and therefore most of the time
only one CPU core is being used.
The \texttt{M0} cross-matching query is CPU-bound on all nodes and
the full materialisation of the large intermediate results into memory 
introduces significant degradations of the performance,
making it unfeasible to process the source lists within  
MeerLICHT's cadence time. 

Smaller scans over sorted partitioned tables and less complex computations
expose better instruction code locality and reduce intermediates
for the \texttt{M5} cross matching,
from which the overall \texttt{M5} pipeline performance distinctly benefits.
The \texttt{M5} cross-match query is about three orders of magnitude faster
and behaves linearly.
The queries in \texttt{M5} utilise the available resources differently,
since the cross-matching requires less CPU time.
All queries are bound to maximum row numbers and 
finish execution within limited amounts of times.
Internal database index structures further speed up the
queries to run sublinear,
giving the pipeline an additional boost of a factor of two.
The partitioned database schema better predicts query behaviour
on the long term, controls the overall pipeline performance and
scales to larger numbers of sources.

\section{Conclusions}
\label{sec:concl}

High-cadence astronomical observatories have the potential 
to building up extremely large databases of catalogued sources and 
their light curves.
Making scientific discoveries with the use of databases
rely on the ability to efficiently grind the massive amounts of data.
In this work, we matched optimised Big Data storage models
to pipeline query access patterns in a layered
storage system. 
The layers present the data in different formats,
going from coarse high-level overviews at the top (all-sky) 
to the fine-grained details at the bottom tier (declination strip).

This work describes a scalable solution 
for the full-source database 
for the MeerLICHT and planned BlackGEM wide-field optical 
telescopes.
We adopted the Transients Pipeline (TraP) 
database schema and pipeline queries from the 
LOFAR Transients Key Science Project 
as a baseline
to ingest, process and store
optical data from binary catalog \texttt{FITS} files.
We investigated the database schemas and query modules to
optimise source cross matching and achieve
long-term sublinear run times for all pipeline queries.
We monitored all queries individually to study their long-term behaviour.
Experiments with real data from the IPHAS Survey showed that
the modified TraP baseline cross-match module is not scalable 
towards optical source densities.
We developed an alternative cross-match module for
database schema and query optimisations
that improved the pipeline runs significantly.
The column-oriented database schema design 
in which the data are partitioned horizontally according to declination strips
allows the data\-base to grow in size and 
simultaneously to run all pipeline queries in 
constant time,
making this the preferred schema for
processing large source lists at high cadence.

Cross-match algorithms of associating a new source list 
with the stored known sources
are highly sensitive to their implementations.
The list of known sources is reduced significantly by 
maintaining an up-to-date compact statistical \emph{sky model}.
After operations start, 
the model's size settles relatively early and 
therefore the cross-matching avoids
large scans over continuously growing tables.
We accelerated the cross-match positional look-ups by three orders of magnitude 
with the use of MonetDB's default hash indexes on a single sorted column of declination values.
The overall pipeline speed was increased by two orders of magnitude.
Most queries run in constant time and only a few run in linear time 
with known upper limits.

In related tests, 
we noticed that the creation of a three-dimensional tree index structure 
of the known cataloged sources becomes unbalanced after a while
and that tree rebuilds are expensive,
which slows down the pipeline run.
This makes the $k$d tree, where $k=3$ for the Cartesian co-ordinates,
not the appropriate index structure
in the sky model build-up 
phase, 
however, in a read-only static database it might well be  
the preferred index structure for fast positional look-ups and cross-matching.
In this context, further Bkd tree performance investigations,
especially in the treatment of updating and removing points from the tree,
optimising memory buffer sizes, number of trees in memory and
tuning access to main memory 
may demonstrate its usefulness in dynamic, distributed, databases.

Performance results showed that the processing of IPHAS data, 
with similar average and peak source densities MeerLICHT
will encounter,
was feasible well within cadence limits of 25 seconds.
Source lists with average densities could be processed at
rates of 5 seconds per image on nodes with
modest CPUs and large RAM.

MeerLICHT's larger field of view implies larger source lists,
but the sublinear-time behaviour and scalability of most queries,
including the cross-matching,
will not affect the performance.
Queries that run in linear time will have similar performance for MeerLICHT,
since the dependence scales with source density and not 
source list size for the most intense queries.
The assumption of 12 hours of observation per night and the 
offline mode that allows delays due to peaks
will relax the criteria, 
meaning that with the alternative database and 
partitioning schema MeerLICHT full-source data can be processed 
at the one minute cadence.

The BlackGEM array will consist of three telescopes, 
where each one will observe a different patch of the sky and 
will produce its own data stream, similar to MeerLICHT is doing.
This simplifies parallelisation at an early stage, where
the fields of view can be processed independently
by multiple threads and database connections working 
on distinct partitions of the data.
The in-memory column-oriented data storage structures of MonetDB
matches with the data ingestion and pipeline-specific queries.
Independent parallel data streams will make the BlackGEM 
full-source pipeline capable of processing 
the data within its cadence time of one minute.

Development of the MonetDB database and the MeerLICHT \& BlackGEM 
full-source pipeline continues,
where we will address improvements on data-partition querying, multi-dimensional tree indexing
and techniques to visualise data in full-source archives.

\section*{Acknowledgement}
      We would like to thank the referee for the comments and 
      suggestions that improved the paper.
      BS and MK acknowledge funding from the 
      Dutch National Science Organisation
      NWO under Grant
      \emph{P4164 NWO project Big Bang Big Data (628-002-004)}

\section*{References}

\onecolumn

\begin{appendix} 

\section{Insertion queries}

\subsection{\texttt{I3}, loading FITS data}
\label{app:i3}

\begin{lstlisting}[language=sqlmore
                  ,basicstyle=\ttfamily\tiny
                  ,keywordstyle=\color{blue}
                  ,stringstyle=\color{green}
                  ,commentstyle=\color{red}
                  ]
/* tabname is the data-vaults table that was generated at the moment
   the FITS file was attached to the database schema. 
   This procedure really loads the data permanently into the database. */
CALL fitsload('%(tabname)s');
\end{lstlisting}

\subsection{\texttt{I4}, loading FITS header data}
\label{app:i4}

\begin{lstlisting}[language=sqlmore
                  ,basicstyle=\ttfamily\tiny
                  ,keywordstyle=\color{blue}
                  ,stringstyle=\color{green}
                  ,commentstyle=\color{red}
                  ]
INSERT INTO image
  (run
  ,extver
  ,irafname
  ,wffpos
  ,wffband
  ,wffpsys
  ,wffid
  ,jd
  ,mjd
  ,magzpt
  ,exptime
  ,airmass
  ,extinct
  ,apcor
  ,percorr
  )
VALUES
  (%(run)s
  ,%(extver)s
  ,'%(irafname)s'
  ,%(wffpos)s
  ,CAST('%(wffband)s' AS CHAR(1))
  ,'%(wffpsys)s'
  ,%(wffid)s
  ,%(jd)s
  ,%(mjd)s
  ,%(magzpt)s
  ,%(exptime)s
  ,%(airmass)s
  ,%(extinct)s
  ,%(apcor)s
  ,%(percorr)s
  )
;
\end{lstlisting}

\subsection{\texttt{I5}, inserting FITS data into permanent table}
\label{app:i5}

\begin{lstlisting}[language=sqlmore
                  ,basicstyle=\ttfamily\tiny
                  ,keywordstyle=\color{blue}
                  ,stringstyle=\color{green}
                  ,commentstyle=\color{red}
                  ]
INSERT INTO extractedsource
  (id
  ,number
  ,isophotal_flux
  ,total_flux
  ,core_flux
  ,x_coordinate
  ,y_coordinate
  ,gaussian_sigma
  ,ellipticity
  ,position_angle
  ,peak_height
  ,areal_1_profile
  ,areal_2_profile
  ,areal_3_profile
  ,areal_4_profile
  ,areal_5_profile
  ,areal_6_profile
  ,areal_7_profile
  ,areal_8_profile
  ,core1_flux
  ,core2_flux
  ,core3_flux
  ,core4_flux
  ,ra
  ,"dec"
  ,classification
  ,statistic
  ,core5_flux
  ,skylev
  ,skyrms
  ,bad_pixels
  ,blank31
  ,blank32
  ,ra_deg
  ,dec_deg
  ,dec_zone_deg
  ,x
  ,y
  ,z
  ,extver
  ,image
  )
SELECT id
      ,number
      ,isophotal_flux
      ,total_flux
      ,core_flux
      ,x_coordinate
      ,y_coordinate
      ,gaussian_sigma
      ,ellipticity
      ,position_angle
      ,peak_height
      ,areal_1_profile
      ,areal_2_profile
      ,areal_3_profile
      ,areal_4_profile
      ,areal_5_profile
      ,areal_6_profile
      ,areal_7_profile
      ,areal_8_profile
      ,core1_flux
      ,core2_flux
      ,core3_flux
      ,core4_flux
      ,ra
      ,"dec"
      ,classification
      ,statistic
      ,core5_flux
      ,skylev
      ,skyrms
      ,bad_pixels
      ,blank31
      ,blank32
      ,ra_deg
      ,dec_deg
      ,dec_zone_deg
      ,x
      ,y
      ,z
      ,extver
      ,%(image_id)s
  FROM tmpextrsrc
;
\end{lstlisting}

\section{Baseline cross-match module \texttt{M0}}
\label{app:q0}

\begin{lstlisting}[language=sqlmore
                  ,basicstyle=\ttfamily\tiny
                  ,keywordstyle=\color{blue}
                  ,stringstyle=\color{green}
                  ,commentstyle=\color{red}
                  ]
DECLARE iassoc_r_arcsec, iassoc_r_deg, idist_const DOUBLE;
SET iassoc_r_arcsec = CAST(%(dr_arcsec)s AS DOUBLE);
SET iassoc_r_deg = iassoc_r_arcsec / 3600;
SET idist_const = PI() * iassoc_r_arcsec / 1296000;

SELECT t0.runcat
      ,t0.xtrsrc
      ,3600 * DEGREES(2 * t0.dist_const) AS distance_arcsec
  FROM (SELECT rc1.id AS runcat
              ,x1.id AS xtrsrc
              ,ASIN(SQRT( (rc1.x - x1.x) * (rc1.x - x1.x)
                        + (rc1.y - x1.y) * (rc1.y - x1.y)
                        + (rc1.z - x1.z) * (rc1.z - x1.z)
                        ) / 2) AS dist_const
          FROM extractedsource x1
              ,image i1
              ,runningcatalog rc1
         WHERE i1.run = irun
           AND x1.image = i.id
           AND rc1.dec_zone_arcsec BETWEEN CAST(FLOOR(3600 * x1.dec_deg - iassoc_r_arcsec) AS INTEGER)
                                       AND CAST(FLOOR(3600 * x1.dec_deg + iassoc_r_arcsec) AS INTEGER)
           AND rc1.ra_deg BETWEEN x1.ra_deg - alpha(x1.dec_deg, iassoc_r_deg)
                              AND x1.ra_deg + alpha(x1.dec_deg, iassoc_r_deg)
       ) t0
 WHERE t0.dist_const < idist_const
;
\end{lstlisting}

\section{Alternative cross-match module \texttt{M5}}
\label{app:M5}

\begin{lstlisting}[language=sqlmore
                  ,basicstyle=\ttfamily\tiny
                  ,keywordstyle=\color{blue}
                  ,stringstyle=\color{green}
                  ,commentstyle=\color{red}
                  ]
/* Create the partition rc_zone table, where the table name
   is appended with the zone id. Then, append it to the MERGE
   table runcat. */
CREATE TABLE "%(rc_zone)s"
  (id INT NOT NULL
  ,xtrsrc INT NOT NULL
  ,datapoints INT NOT NULL
  ,active BOOLEAN NOT NULL DEFAULT TRUE
  ,avg_ra DOUBLE PRECISION NOT NULL
  ,avg_dec DOUBLE PRECISION NOT NULL
  ,avg_ra_deg DOUBLE PRECISION NOT NULL
  ,avg_dec_deg DOUBLE PRECISION NOT NULL
  ,avg_dec_zone_deg TINYINT NOT NULL
  ,x DOUBLE PRECISION NOT NULL
  ,y DOUBLE PRECISION NOT NULL
  ,z DOUBLE PRECISION NOT NULL
  )
;
ALTER TABLE runcat ADD TABLE "%(rc_zone)s";

/* Declare the temporary new zoned runcat table rcz */
CREATE SEQUENCE "rcz_seq" AS INT START WITH "%(rcz_seq_start)s";
DECLARE TABLE rcz
  (id INT NOT NULL DEFAULT NEXT VALUE FOR "rcz_seq"
  ,xtrsrc INT NOT NULL
  ,datapoints INT NOT NULL
  ,active BOOLEAN NOT NULL DEFAULT TRUE
  ,avg_ra DOUBLE NOT NULL
  ,avg_dec DOUBLE NOT NULL
  ,avg_ra_deg DOUBLE NOT NULL
  ,avg_dec_deg DOUBLE NOT NULL
  ,avg_dec_zone_deg TINYINT NOT NULL
  ,x DOUBLE PRECISION NOT NULL
  ,y DOUBLE PRECISION NOT NULL
  ,z DOUBLE PRECISION NOT NULL
  )
;

/* rcz is loaded with use of a select statement (for brevity not 
   explicitly shown) that unions all the relevant rc_zone tables. */
INSERT INTO rcz
      (id, xtrsrc, datapoints, avg_ra,avg_dec
      ,avg_ra_deg,avg_dec_deg,avg_dec_zone_deg
      ,x,y,z)
"%(select_query)s"
;

/* The cross-match query. Python variables are defined analogously to M2 */
SELECT t1.runcat
      ,t1.xtrsrc
      ,3600 * DEGREES(2 * ASIN(SQRT(t1.dist) / 2)) AS distance_arcsec
  FROM (SELECT z0.id AS runcat
              ,t0.id AS xtrsrc
              ,  (z0.x - t0.x) * (z0.x - t0.x)
               + (z0.y - t0.y) * (z0.y - t0.y)
               + (z0.z - t0.z) * (z0.z - t0.z) AS dist
          FROM rcz z0
              ,(SELECT id
                      ,dec_deg - "%(iradius)s" AS decmin
                      ,dec_deg + "%(iradius)s" AS decmax
                      ,ra_deg - alpha(dec_deg, "%(iradius)s)" AS ramin
                      ,ra_deg + alpha(dec_deg, "%(iradius)s)" AS ramax
                      ,x
                      ,y
                      ,z
                  FROM tmpextrsrc x0
               ) t0
         WHERE z0.avg_dec_deg BETWEEN t0.decmin AND t0.decmax
           AND z0.avg_ra_deg BETWEEN t0.ramin AND t0.ramax
       ) t1
      ,rcz z1
      ,tmpextrsrc x1
 WHERE z1.id = t1.runcat
   AND t1.dist < "%(isint2)s"
   AND x1.id = t1.xtrsrc
;
\end{lstlisting}

\end{appendix}


\begin{thebibliography}{00}

\bibitem[Abadi et~al.(2008)]{Abadi08}
Abadi, D.K., Madden, S.R.\ and Hachem, N., Column-stores vs.\ row-stores: how different are they really? 
In Proceedings of the 2008 ACM SIGMOD international conference on 
Management of data, pages 967–980. ACM, 2008

\bibitem[Abadi~et~al.(2012)]{Abadi12} 
Abadi, D., Boncz, P., Harizopoulos, S., Idreos, S.\ and Madden, S.,
The Design and Implementation of Modern Column-Oriented Database Systems,
Foundations and Trends in Databases, Vol.5, No.3, 2012

\bibitem[Abell et~al.(2009)]{Abell} Abell et al., 
2009, LSST Science Collaborations, arXiv, arXiv:0912.0201L

\bibitem[Alam et~al.(2015)]{Alam}
Alam, S. et al., 2015, ApJS, 219, 12A

\bibitem[Barentsen et~al.(2014)]{Barentsen}
Barentsen, G., Farnhill, H.J., Drew, J.E., Gonz\'alez-Solares, E.A.\ et al., 2014,
MNRAS, 444, 3230B

\bibitem[Becla and Wang(2014)]{Becla} Becla, J.\ \& Wang, D.L., 
Enabling Scalable Data Analytics for LSST and Beyond, 
Exascale Radio Astronomy, 
AAS Topical Conference Series Vol. 2. 
Proceedings of the conference held 30 March--4 April, 2014 in Monterey, California. 
Bulletin of the American Astronomical Society, Vol. 46, \#3, \#303.03, 
2014era, conf, 30303B

\bibitem[Bloemen et~al.(2015)]{Bloemen}
Bloemen, S., Groot, P., Nelemans, G. and Klein-Wolt, M.,
2015, ASPC, 496, 254B

\bibitem[Boncz et al.(1999)]{BMK99}
Boncz, P.A., Manegold, S.\ and Kersten, M.L., Database
architecture optimized for the new bottleneck: Memory access. In
VLDB, volume 99, pages 54–65, 1999.

\bibitem[Boncz(2002)]{Boncz}
Boncz, P.A., Monet: A Next-Generation DBMS Kernel For 
Query-Intensive Applications. 
Ph.D.\ thesis, Universiteit van Amsterdam, May 2002.

\bibitem[Brederode et al.(2016)]{Brederode}
Brederode, L.R., van den Heever, L., Esterhuyse, W.~and Jonas, J.L.,
2016, SPIE, 9906E, 25B

\bibitem[Broekema~et~al.(2012)]{Broekema} Broekema, P.C., Van Nieuwpoort, R.V. \& Bal, H.E., 
ExaScale High Performance Computing in the Square Kilometer Array, 
Proceedings of the 2012 workshop on High-Performance Computing for Astronomy 
(AstroHPC'12), Delft, the Netherlands, June, 2012

\bibitem[Cordier~et~al.(2015)]{SVOMGWAC}
Cordier, B., Wei, J., Atteia, J.-L., Basa, S., Claret, A., Daigne, F., Deng, J., 
Dong, Y., Godet, O., Goldwurm, A., G\"otz, D., Han, X., Klotz, A., Lachaud, C., 
Osborne, J., Qiu, Y., Schanne, S., Wu, B., Wang, J., Wu, C., Xin, L., Zhang, B. and Zhang, S.-N.,
2015, PoS (SWIFT~10) 005

\bibitem[Gray et~al.(2006)]{Gray}
Gray J., Nieto-Santisteban M.A., Szalay A. 2006, The Zones Algorithm
for Finding Points-Near-a-Point or Cross-Matching Spatial Datasets, 
Microsoft, Johns Hopkins University, MSR TR 2006 52

\bibitem[H\'eman(2015)]{Heman} 
Updating Compressed Column-Stores, 
Ph.D.\ thesis, Vrije Universiteit Amsterdam, October 2015.

\bibitem[Hennessy \& Patterson(2012)]{HP12} 
Hennessy, J. and Patterson, D., Computer architecture: a quantitative approach (5th edition). 
Morgan Kaufmann Publishers Inc., 2012

\bibitem[Hodapp et~al.(2004)]{Hodapp} 
Hodapp, K.W., et al., 2004, AN, 325, 636

\bibitem[Ivanova et~al.(2007)]{Ivanova}
M. Ivanova, N. Nes, R. Goncalves, M. Kersten, "MonetDB/SQL Meets SkyServer: 
the Challenges of a Scientific Database", SSDBM, 2007, pp. 13 


\bibitem[Ivanova et~al.(2013)]{Ivanova13}
M. Ivanova, Y. Karg{\i}n, M. Kersten, S. Manegold, Y. Zhang, M. Datcu, D. Molina
"Data Vaults: a Database Welcome to Scientific File Repositories",
SSDBM Baltimore, 2013, MD, USA

\bibitem[Ivezi\'c et~al.(2017)]{Ivezic}
Ivezi\'c, \v Z., Connolly, A.J., Juri\'c, M., arXiv, arXiv:1612.04772

\bibitem[Juric~et~al.(2015)]{Juric} Juric, M., Tyson, T., 2015,
LSST Data Management: Entering the Era of Petascale Optical Astronomy, 
HiA, 16, 675J

\bibitem[Keller et~al.(2007)]{Keller}
Keller, S. C., Schmidt, B. P., Bessell, M. S., Conroy, P. G.,
Francis, P., Granlund, A., Kowald, E., Oates, A. P.,
Martin-Jones, T., Preston, T., Tisserand, P., Vaccarella, A., \&
Waterson, M. F. 2007, Publications of the Astronomical
Society of Australia, 24, 1

\bibitem[Lazio~et~al.(2014)]{Lazio} Lazio, J. W., Kimball, A., Barger, A.J., Brandt, W.N., 
Chatterjee, S., Clarke, T.E., Condon, J.J., Dickman, R.L., Hunyh, M.T., Jarvis, M.J., 
Juric, M., Kassim, N.E., Myers, S.T., Nissanke, S., Osten, R., Zauderer, B. A., 2014,
Radio Astronomy in LSST Era, 
PASP, 126, 196L

\bibitem[Lonsdale et~al.(2009)]{Lonsdale}
Lonsdale, C.J., et al., 2009, arXiv, arXiv:0903.1828

\bibitem[Manegold et~al.(2002)]{Stefan}
Manegold, S., Boncz, P.\ and Kersten, M.L., Generic Database Cost Models for Hierarchical Memory Systems. 
In VLDB, 2002

\bibitem[Murphy~et~al.(2013)]{Murphy} Murphy, T., Chatterjee, S., Kaplan, D.L., Banyer, J., 
Bell, M.E., Bignall, H.E., Bower, G.C., Cameron, R.A., Coward, D.M., Cordes, J. et al., 2013, 
VAST: An ASKAP Survey for Variables and Slow Transients, 
PASA, 30, 6M

\bibitem[Sidirourgos \& Kersten(2013)]{Lefteris}
Sidirourgos, L.\ and Kersten, M.L., Column imprints: a secondary 
index structure. In Proceedings of the ACM SIGMOD Conference 
on Management of Data, pages 893–904, 2013.

\bibitem[Swinbank et~al.(2015)]{Swinbank}
John D. Swinbank, Tim D. Staley, Gijs J. Molenaar, Evert Rol, Antonia Rowlinson, Bart Scheers, Hanno Spreeuw, Martin E. Bell, Jess W. Broderick, Dario Carbone, Alexander J. van der Horst, Casey J. Law, Michael Wise, Rene P. Breton, Yvette Cendes, Stéphane Corbel, Jochen Eislöffel, Heino Falcke, Rob Fender, Jean-Mathias Greißmeier, Jason W.T. Hessels, Benjamin W. Stappers, Adam J. Stewart, Ralph A.M.J. Wijers, Rudy Wijnands, Philippe Zarka
The LOFAR Transients Pipeline,
2015, A\&C, 11, 25S

\bibitem[Szalay \& Blakeley(2009)]{SzBl} 
Szalay, A.S. \& Blakeley, J.A., 2009, The Fourth Paradigm. Data-Intensive
Scientific Discovery, Eds. Hey, T., Tansley, S. \& Tolle, K., Microsoft
Corporation, USAi, p. 5

\bibitem[Tingay et~al.(2013)]{Tingay} Tingay, S.J., Goeke, R. et al., 2013, 
The Murchison Widefield Array: the Square Kilometre Array Precursor at low radio frequencies, 
PASA, 30, 7T

\bibitem[van Haarlem et~al.(2012)]{vHW} van Haarlem, M.P., Wise, M.W. et al., 2012,
LOFAR: The LOw-Frequency ARray
A\&A, 556A, 2V

\bibitem[York et~al.(2000)]{York}
York, D.G. et al., 2000, AJ 120, 1579

\bibitem[Welch et~al.(2009)]{Welch}
Welch, J., et al., 2009, arXiv, arXiv:0904.0762

\bibitem[Zukowski(2005)]{Marcin}
Zukowski, M., Hardware-Conscious DBMS Architecture for Data-Intensive Applications,
Proceedings of the 31st VLDB Conference, Trondheim, Norway, 2005

\end{thebibliography}
\end{document}